\documentclass[11pt]{article}
\pdfoutput=1

\usepackage{jheppub}
\usepackage[T1]{fontenc}

\usepackage{jheppub}
\usepackage{url,comment}
\usepackage{times}
\usepackage{latexsym}
\usepackage{graphicx, graphics, hyperref, amsmath, amssymb, slashed, color, bbm,amsthm, array}
\usepackage[usenames,dvipsnames]{xcolor}
 \usepackage{subfigure}
\usepackage{mdwlist}
\usepackage{multirow}
\usepackage[e]{esvect}

\usepackage{cancel}
\usepackage{pstricks}
\usepackage[toc,page]{appendix}

\newcommand{\hc}{{\rm h.c.}}

\graphicspath{{figs/}}

\DeclareMathOperator{\Tr}{Tr}

\title{Enhanced Higgs Mass in Compact Supersymmetry}
\abstract{The current LHC results make weak scale supersymmetry difficult due to relatively heavy mass of the discovered Higgs boson and the null results of new particle searches.  
Geometrical supersymmetry breaking from extra dimensions, Scherk-Schwarz mechanism, is possible to accommodate such situations. A concrete example, the Compact Supersymmetry model, has   a compressed spectrum ameliorating the LHC bounds  and large mixing in the top and scalar top quark sector with $|A_t|\sim 2m_{\tilde{t}}$ which radiatively raises the Higgs mass. While the zero mode  contribution of the model has been considered, in this paper we calculate  the Kaluza-Klein tower effect to the Higgs mass. Although such contributions are naively expected to be as small as a percent level for 10~TeV Kaluza-Klein modes,
we find the effect significantly enhances  the radiative correction to the Higgs quartic coupling  by from 10 to 50~\%. 
This is mainly because the top quark wave function is pushed out from the brane, which makes the top mass  depend on higher powers in the Higgs field.
As a result  the Higgs mass is enhanced up to 15~GeV from the previous calculation.  We also show the whole parameter space is testable at the LHC run II.  
 }

\author[a,b,c]{Kohsaku Tobioka, }
\emailAdd{tobioka@post.kek.jp}
\affiliation[a]{Raymond and Beverly Sackler School of Physics and Astronomy, Tel-Aviv University, Tel-Aviv 69978, Israel}
\affiliation[b]{Department of Particle Physics and Astrophysics,
Weizmann Institute of Science, Herzl St 234, Rehovot 7610001, Israel}

\author[c,d,e]{Ryuichiro Kitano, }
\affiliation[c]{Theory Center, High Energy Accelerator Research Organization (KEK), 1-1 Oho, Tsukuba 305-0801, Japan}
\affiliation[d]{The Graduate University for Advanced Studies (Sokendai), 1-1 Oho, Tsukuba 305-0801, Japan}
\emailAdd{Ryuichiro.Kitano@kek.jp}

\author[e,f,g]{Hitoshi Murayama, }
\emailAdd{hitoshi.murayama@ipmu.jp}
\affiliation[e]{Kavli Institute for the Physics and Mathematics of the Universe (WPI), University of Tokyo Institutes for Advanced Study, University of Tokyo, 5-1-5 Kashiwanoha,
Kashiwa 277-8583, Japan}
\affiliation[f]{Department of Physics, University of California, Berkeley, 366 LeConte Hall, Berkeley, California 94720, USA}
\affiliation[g]{Theoretical Physics Group, Lawrence Berkeley National Laboratory, 1 Cyclotron Rd, Berkeley, California 94720, USA}

\date{\today}

\preprint{IPMU15-0193. KEK-TH-1873. UCB-PTH-15/12. }

\begin{document}
\maketitle

\section{Introduction}
Supersymmetry is the prime candidate for the physics beyond the standard model (for a review~\cite{Martin:1997ns}). 
It can stabilizes the large hierarchy between the electroweak scale and some high energy scale such as the quantum gravity scale, and also it leads to dynamical electroweak symmetry breaking by radiative correction effects.  
The minimal supersymmetric standard model (MSSM) has been studied as an attractive and minimal model of supersymmetry. 
However under its constrained framework,  the lightest Higgs mass  has an upper bound of  $m_Z\simeq91~\rm GeV$ at tree level, and hence this has a very strong tension with  the recently discovered Higgs boson  at the Large Hadron Collider (LHC) \cite{Aad:2012tfa, Chatrchyan:2012xdj}. As the latest result, the  combined analysis by the ATLAS and CMS says that the measured mass is $125.09\pm 0.21\pm 0.11 \rm GeV$ \cite{Aad:2015zhl}.  To explain the discrepancy it is known that the Higgs mass  can be raised beyond  $m_Z$ by  radiative corrections through large scalar top (stop) mass or  large mixing between left and right stops  (large $A$-term)  but in most cases fine-tuning of less than percent level is accompanied with those radiative corrections \cite{Hall:2011aa}.
Various extensions to the minimal model are also considered in order to accommodate the Higgs mass. One direction is to introduce singlet(s) to the Higgs sector resulting in non-decoupling $F$-term as in NMSSM~\cite{Espinosa:1991gr, Nomura:2005rk, Dine:2007xi, Ellwanger:2009dp} and Dirac-NMSSM~\cite{Lu:2013cta}, or to introduce an extra gauge group under which Higgs is charged resulting in non-decoupling $D$-term  \cite{Batra:2003nj, Maloney:2004rc}. Also a strongly coupled Higgs sector is another possibility \cite{Harnik:2003rs, Kitano:2012cz, Kitano:2012wv}.  

On the other hand, supersymmetric particles (sparticles) has been  extensively searched for at the LHC. So far the null result has been found at the LHC run I and therefore the strong constraints on the parameter space are obtained. Typical supersymmetric models such as the Constrained MSSM (CMSSM) \cite{Hall:1983iz} has an exclusion bound on sparticles mass beyond TeV \cite{Aad:2014wea}, which is another tension in addition to the discrepancy of the Higgs mass.  The sparticle searches are mainly based on missing transverse energy, $\slashed{E}_T$,~\footnote{To be precise, this should be called as missing transverse momentum because we can only measure a missing quantity  constructed by transverse momentum conservation, $\slashed{\vec{P}}_T =-\sum \vec{P}_T^{\rm vis}$. In fact, in a scenario of compressed spectrum, the energy carried out of the detector is large but the missing  momentum can be small. However here we use the convention that missing transverse momentum is called as $\slashed{E}_T$.} 
 motivated by $R$-parity, and the corresponding exclusion bounds are often strong but  still model dependent. 
 The signal becomes weaker, even in presence of enouch sparticle production, 
 due to  small missing energy  when a  mass spectrum is compressed~\cite{LeCompte:2011fh, Murayama:2012jh} or sparticles decay to new states~\cite{Fan:2011yu, Fan:2012jf},  or due to lack of missing energy signal  when  $R$-parity is violated~\cite{Barbier:2004ez, Csaki:2011ge, Ruderman:2012jd, Bhattacherjee:2013gr, Csaki:2013jza}. 

The Compact Supersymmetry model has a possibility to accommodate those tensions because it has a compressed spectrum and large  $A$-term~\cite{Murayama:2012jh}. 
The model is embedded in 5D spacetime with a simple extra dimension, $S^1/{\mathbb{Z}}_2$, and has a field configuration that quark, lepton and gauge superfields are in the bulk while Higgs fields are localized on a brane. 
Supersymmetry  is broken by  non-trivial boundary conditions of the extra dimension, called the Scherk-Schwartz mechanism \cite{Scherk:1978ta, Scherk:1979zr}.  As a direct consequence of the field configuration and the Scherk-Schwartz mechanism, the universal soft masses, which is important for a compressed spectrum,  and near maximal mixing by $A$-term ($|A_t|\sim 2m_{\tilde{t}}$), which enhances the Higgs mass beyond $m_Z$, are obtained. Requirement of the successful electroweak symmetry breaking fixes  the supersymmetric Higgs mass, $\mu$ term, which determines the mass scale of the lightest sparticle (LSP). 
As a result,  a spectrum with mass compression of a few hundreds of GeV is  realized in generic parameter space ameliorating the LHC bounds. 
The model has further attractive features. The geometrical nature of supersymmetry breaking does not introduce the conventional CP or flavor problems, and there are only three free parameters two of which are left after requiring the  successful electroweak symmetry breaking. 

Regarding the Higgs mass, the previous computation for the model has included the only zero mode contributions, that is the MSSM contribution, and  the Higgs mass was expected to be 119 to 125~GeV even for TeV sparticle thanks to the large $A$-term. 
Interestingly, without adding any extra things such as singlets, it can be further enhanced by a contribution from the Kaluza-Klein (KK) tower.  This possibility is implied by the Constrained Standard model~\cite{Barbieri:2000vh} which is another model with the Scherk-Schwarz mechanism and has a similar Higgs sector of the Compact Supersymmetry model. 
It is pointed out in Ref.~\cite{Barbieri:2000vh} that  the KK tower drastically changes the Higgs potential  and  the Higgs mass is highly enhanced even with very light stop. While the Constrained Standard model  is not compatible with the LHC results, the KK effect remains interesting especially because the observed Higgs mass of 125 GeV  is relatively heavy for the MSSM.  
In this paper we revisit the effect of KK tower to the Higgs sector with general supersymmetry breaking parameter because the Constrained Standard model focused on the maximal breaking case. Then we apply the result to the Higgs mass calculation. We find the effect remains large even when the KK modes are at $\cal O$(10) TeV, 
and the enhancement of the Higgs mass from the MSSM calculation is from 5 to 15~GeV in interesting parameter regions. 
Since $125~\rm GeV$ Higgs mass is realized in a lower supersymmetry breaking scale, 
 TeV range of sparticle mass is still compatible with the observed Higgs mass and  the LHC sparticle searches in this model. An interesting upper bound of sparticle mass is obtained by the dark matter relic density, which implies that the whole parameter space can be tested by the LHC run II.

The Scherk-Schwarz mechanism has been discussed with a special attention to the UV-finite feature: not only quadratic divergence is absent but also log divergence is absent \cite{Antoniadis:1998sd, Delgado:1998qr, Barbieri:2000vh, ArkaniHamed:2001mi, Ghilencea:2001ug, Delgado:2001ex, Contino:2001gz, Barbieri:2001dm, Masiero:2001im, Delgado:2001xr, Kim:2001gk}. This is because of the non-local nature of supersymmetry breaking. We see such a feature in our calculation of the radiative corrections from the KK tower.  
Also many models beyond the Standard Model with various field configurations are discussed in Refs~\cite{Mirabelli:1997aj, Pomarol:1998sd, Antoniadis:1998sd, Delgado:1998qr,  Chacko:2000fn, Barbieri:2000vh, ArkaniHamed:2001mi, Barbieri:2001dm, Barbieri:2001yz, Delgado:1999sv, Delgado:2001si, Delgado:2001ex, Delgado:2001xr, Barbieri:2002uk, Barbieri:2003kn, Kitano:2006gv, Dimopoulos:2014aua}.  
As a further extension, the mechanism is used in Folded Supersymmetry models \cite{Burdman:2006tz, Cohen:2015gaa}.

In the following section we present an overview of the Scherk-Schwartz mechanism and the Compact Supersymmetry model. We compute radiative corrections of the KK tower to the Higgs sector in Sec.~\ref{sec:Effpotential} and apply the result to the Higgs mass calculation in Sec.~\ref{sec:Higgsmass}. In Sec.~\ref{sec:Expbound}. we study experimental bounds and show the LHC bound is certainly weaker due to the compressed spectrum. 
We conclude in Sec.~\ref{sec:conclude}. 

\section{Overview}
 \subsection{Scherk-Schwartz Mechanism and  Compact Supersymmetry Model}
We consider a single compact extra dimension with its coordinate $y$ identified by translation, ${\cal T}:y\to y+2\pi R$, and reflection, ${\cal Z}: y\to -y$, where $R$ is the radius of the extra dimension. This is $S^1/{\mathbb{Z}}_2$ orbifold, and it is subject to two consistency conditions, 
	\begin{align}
	{\cal P}^2=1 \ , \quad{\cal  PTP=T}^{-1} \ . 
	\end{align}
  The minimum supersymmetry in 5D  corresponds to ${\cal N}=2$ in 4D, leading to a global $SU(2)_R$ symmetry.  In presence of such a global symmetry, $SU(2)_R$ doublets, such as gauginos, can have non-trivial boundary conditions with a {\it twist} of $\alpha$ under the translation,  
	\begin{align}
	{\cal P}=\left(\begin{array}{c c}1 &0 \\  
	0 &-1 \end{array}\right)
	, \quad 
	{\cal T}=e^{(2\pi \alpha)i  \sigma_2}=
	\left(\begin{array}{c c}\cos(2\pi\alpha) &\sin(2\pi\alpha) \\  
	-\sin(2\pi\alpha) &\cos(2\pi\alpha) \end{array}\right),   \label{eq:PT}
	\end{align}
where $0\leq\alpha\leq1/2$. We are interested in $\alpha\ll1$. On the other hand, $SU(2)_R$ singlets have only trivial boundary conditions of ${\cal P}=\pm 1$ and ${\cal T}=1$ because they do not have a matrix structure. The finite twist of $\alpha$ leads to mass splitting in supermultiplets and then supersymmetry is broken. This mechanism is called the Scherk-Schwarz mechanism \cite{Scherk:1978ta, Scherk:1979zr}. 
\begin{figure*}[t]
 \begin{center}
  \includegraphics[width=0.55\linewidth]{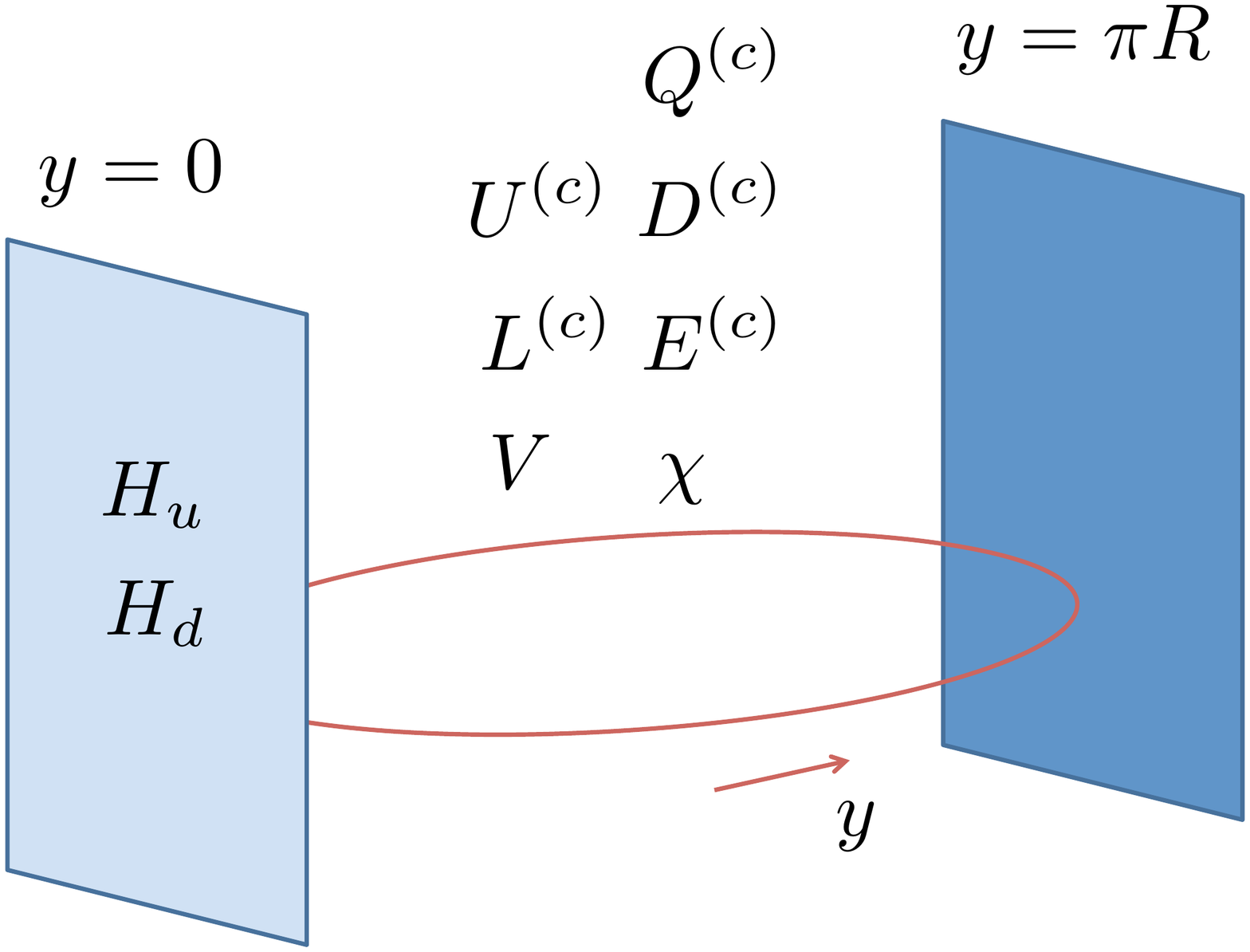}
\vspace{-8pt}
 \end{center}
 \caption{\label{fig:Config}
  Field configurations of the Compact Supersymmetry model. Hypermultiplets of quark and lepton and gauge fields are living in the whole 5D spacetime while  Higgs chiral superfields are localized on a brane. }
\end{figure*}

The Scherk-Schwarz mechanism is applied in the Compact Supersymmetry model \cite{Murayama:2012jh} in which three-generation of quarks and leptons as well as the Standard Model gauge groups are embedded in a bulk (extra dimension) and two Higgs fields, $H_u$ and $H_d$,  are localized on a brane at $y=0$.  A schematic picture of the model seen in Fig.~\ref{fig:Config}. A bulk field forms a hypermultiplet that have two chiral superfields.  One of them, $\Phi=Q, U, D, L,  E$,  has a reflection property of ${\cal P}=1$ and keeps zero mode. The other chiral superfield is denoted with a superscript  as $\Phi^c$ and has  a property of ${\cal P}=-1$ leading to absence of zero mode.  Under the translation,  the squark (slepton) in $\Phi$ mixes with the squark (slepton) in $\Phi^c$ with the same twist as in Eq.~\eqref{eq:PT} because squarks (sleptons) are $SU(2)_R$ doublets.  
Since the twist parameter is unique in a single global symmetry, $SU(2)_R$ doublets, gauginos, squarks, sleptons and gravitinos, must have the same soft mass of $\alpha/R$ after the KK expansion. 
A similar model where Higgs lives in the bulk as well as matters is studied in Ref.~\cite{Barbieri:2001yz}. 

Using ${\cal P}=1$ chiral superfields, we can write Yukawa couplings and $\mu$ term, 
	\begin{align}
	{\cal W}_{\rm brane}= \delta(y) \left\{y_{U5} H_u Q U+ y_{D5}H_d QD +y_{L5}H_d LE + \mu H_uH_d  \right\} . 
	\end{align}
 Matter fields have a non-trivial twist by the Scherk-Schwarz mechanism giving trilinear scalar couplings which corresponds to soft supersymmetry breaking of Yukawa couplings, $A$-terms. However, the Higgs are just 4D field and do not feel supersymmetry breaking at the tree level, and therefore soft terms related only to Higgs are absent. In summary, using the conventional MSSM notation, the soft breaking terms are given by
	\begin{align}
	M_{1/2}=\frac{\alpha}{R} , \ \ m^2_{\tilde{Q}, \tilde{U}, \tilde{D}, \tilde{L}, \tilde{E}}=\frac{\alpha^2}{R^2}, 
	\ \ A_0=-\frac{2\alpha}{R},   \label{eq:softterms}
	\ \ m_{H_u, H_d}^2=0, \ b=0. 
	\end{align}
This is more easily  derived in an equivalent picture of the Radion Mediation in Sec.~\ref{sec:RadionMed}. We emphasize that the model realizes the large $A$-term, $A_0\approx -2 m_{\tilde{t}}$,  which enhances the lightest Higgs mass and may explain the observed value, and we will show that it is enhanced even beyond this expectation. 
It is noteworthy that the conventional  supersymmetric CP and flavor problems are absent thanks to geometric nature of supersymmetry breaking. 
Then model has a compact parameter set of $\alpha, R$, and $\mu$ even more constrained than the CMSSM, which implies testability of the model. Note that Ref.~\cite{Kitano:2006gv} addresses the same framework with the soft breaking of Eq.~\eqref{eq:softterms}  as a possible solution to the little hierarchy problem \cite{Barbieri:2000gf}. %

\subsection{Picture of Radion Mediation}\label{sec:RadionMed}
On $S^1/{\mathbb{Z}}_2$ orbifold, Refs.~\cite{Marti:2001iw, Kaplan:2001cg}~ show that the Scherk-Schwarz mechanism is equivalent to the Radion Mediation \cite{Chacko:2000fn}. The Radion Mediation gives supersymmetry breaking by the $F$-term vacuum expectation value of the Radion chiral superfield, $T$, and hence it is more comprehensive. The 5D bulk action based on the 4D superspace \cite{ArkaniHamed:2001tb}  is 
	\begin{align}
	{\cal K}_5=\ &
 	\frac{T+T^\dag}{2R}\left\{
	{\Phi^\dag} e^{-V}{\Phi} +{\Phi^c} e^{V}{\Phi^{c\dag}} 
	\right\}  \nonumber\\
	+&\frac{1}{8kg_5^2}\frac{2R}{T+T^\dag}
	\Tr\! \left[ (\partial_5+\sqrt{2}{\chi}^{\dag})e^{-V} (-\partial_5+\sqrt{2}{\chi})e^{V} 
	+\frac{\partial_5 e^{-V}\partial_5 e^{V}}{2}+(\chi \chi+{\chi}^{\dag}{\chi}^{\dag})  \right]
	, 
	\\
	{\cal W}_5=\ & \Phi^c (\partial_5  -\sqrt{2}  \chi) \Phi 
	+\frac{1}{16 k g_5^2 }\frac{T}{R}\Tr[ { W_\alpha} { W^\alpha}] \ , \label{eq:W5}
	\end{align}
where 
$V$ is gauge superfield, $\chi$ is adjoint chiral superfield, and  ${\rm Tr} [T^aT^b] =k\delta^{ab}$.  
$V$ and $\chi$  form a 5D vector supermultplet.  
Here, fields have trivial boundary conditions under translation, ${\cal T}=1$, but supersymmetry breaking is introduced by  the Radion VEV,  
	\begin{eqnarray}
	\langle T \rangle=R+ \theta^2 F_T\ . 
	\end{eqnarray}
As shown in Ref.~\cite{Kaplan:2001cg}, $F_T$ can be removed from the action if a twist is introduced in the global $SU(2)_R$ space, resulting in a correspondence of $F_T =2 \alpha$.  Then the gaugino mass is $M_{1/2}=F_T/(2R)=\alpha/R$ from Eq.\eqref{eq:W5}. 
 
It is easy to see the size of other soft terms.  Since K\"ahler potential for matter fields is not canonically normalized, we normalize the bulk fields by shifting $F$ terms of matter fields such that  
	\begin{align}
	\{Q^{(c)}, U^{(c)}, D^{(c)}, L^{(c)}, E^{(c)} \} \to \left(1-\frac{\alpha}{R}\theta^2 \right)
	\{Q^{(c)}, U^{(c)}, D^{(c)}, L^{(c)}, E^{(c)} \} .
	\label{eq:redefinition}
	\end{align}
The squark and slepton masses are given by residual $(\alpha/R)^2\theta^2\bar\theta^2$ term. Also, a Yukawa coupling comes up with a large $A$-term, for example, the field redefinition of Eq.~\eqref{eq:redefinition} leads to 
	\begin{align}
	&\int\! d^2\theta \ y_{U5} H_u Q U \to \int\! d^2\theta \ \left(1-\frac{\alpha}{R}\theta^2 \right)^2 y_{U5} H_u Q U 
	=\int\! d^2\theta \ \left(1-\frac{2\alpha}{R}\theta^2 \right) y_{U5} H_u Q U \ .
	\end{align}
$A_0=-2\alpha/R$ is shown. It is easy to see that this supersymmetry breaking has the minimal flavor violation structure. 
Therefore all the soft terms  in the Radion Mediation matches with Eq.~\eqref{eq:softterms}.

\section{Effective Potential}\label{sec:Effpotential}
We calculate the effective potential from 1-loop of all the KK modes of top quark and squark to take into account their effect to the lightest Higgs mass. The 5D Lagrangian for the up-type squark and quark bilinears  is given by
	\begin{align}
	{\cal L}_5=\ & \tilde{Q}^\dag (\partial^2 -\partial^2_5)\tilde{Q} +\tilde{Q}^{c\dag} (\partial^2 -\partial^2_5)\tilde{Q}^c
	+\tilde{U}^\dag (\partial^2 -\partial^2_5)\tilde{U} +\tilde{U}^{c\dag} (\partial^2 -\partial^2_5)\tilde{U}^c
	\nonumber\\
	&
	+\delta(y) \left( y_{U5} H_u \tilde{Q}\partial_5 \tilde{U}^{c*}+ y_{U5} H_u \tilde{U}\partial_5 \tilde{Q}^{c*}+\hc \right)
	-|\delta(y) y_{U5} H_u \tilde{Q} |^2	-|y_{U5} H_u \tilde{U} \delta(y)|^2 \ 
	\nonumber\\
	&
	+\overline{\Psi}_{Q}(i\slashed{\partial} +\gamma_5 \partial_5)\Psi_Q
	+\overline{\Psi}_U(i\slashed{\partial} +\gamma_5 \partial_5)\Psi_U
	-\delta(y) y_{U5}{H_u}\overline{\Psi}^c_{Q}P_L\Psi_{U}+\delta(y)y_{U5}^*\overline{\Psi}_{U}P_R\Psi_{Q}^c H_u^*
	,\end{align}
where $\tilde{Q}^{(c)}=\tilde{Q}^{(c)}(x,y), \tilde{U}^{(c)}=\tilde{U}^{(c)}(x,y), H_u=H_u(x)$, and $\Psi$ represents a 4-component fermion,   
	\begin{eqnarray}
	\Psi_Q\equiv \left(\begin{array}{c}{Q}(x,y) \\  \overline{Q}^c(x,y)\end{array}\right), \quad
	\Psi_U\equiv \left(\begin{array}{c}{U}(x,y) \\  \overline{U}^c(x,y)\end{array}\right). 
	\end{eqnarray}

Before going to the effective potential, we have to obtain Higgs-dependent mass eigenvalues. 
Since it is difficult to obtain the mass spectrum after the KK expansion due to infinite mixing terms coming from $\delta(y)$, we derive mass eigenvalues by solving 5D equations of motion. We first generally solve Dirac and Klein-Gordon equations in the bulk respecting properties under the reflection, secondly constraint the solutions by  the boundary condition of the translation (the Scherk-Schwarz mechanism),  and 
finally determine the coefficients by integrating around the brane at $y=0$. 
The integration around the brane gives the Higgs field dependence to quark and squark masses.  The detail is given in Appendix~\ref{app:spectrum}. 

The above computation leads to consistency conditions of mass in Eqs.~(\ref{eq:mfcondition}, \ref{eq:mbcondition}). One  for top quark is  
	\begin{eqnarray}
	\tan^2(M_F \pi R) = \left(\frac{y_{t5} H_u}{2}\right)^2 \ , 
	\end{eqnarray}
and this gives mass eigenvalues of KK tower in presence of Higgs VEV, 
	\begin{align}
	M_F=\frac{n}{R}\pm M_t(H_u)   \quad  (n:\rm integer) \ ,
	\end{align}
where
	\begin{eqnarray} 
	 M_t(H_u) \equiv \frac{1}{\pi R}\arctan\left(\frac{y_{t5} H_u}{2}\right)
	 =y_t H_u- \frac{(\pi R)^2( y_t H_u)^3}{3}+{\cal O}(H_u^5 R^4)
	  .  \label{topmass}
	\end{eqnarray}
The top Yukawa coupling in 4D is given by $y_t=y_{t5}/2\pi R$. Note that the top mass, $M_t$, is not only proportional to $H_u$ but also has higher powers of $H_u$. This can be understood as follows.   
The wave function of top quark zero mode is flat in absence of the Higgs VEV, but is distorted by the non-zero Higgs VEV at $y=0$. Especially, top quark tends to reduce overlap with Higgs to minimize the energy. This is why top mass has non-trivial dependence of $H_u$. In the language of 4D effective theory, 
 the term of ${\cal O}(H_u^3)$  is due to higher dimensional operators generated by the non-zero KK modes and the coefficient actually could be explained by summation  of the tower, $\sum_{n=1}^\infty \frac{1}{n^2/R^2}=R^2\zeta(2)={\pi^2 R^2}/{6}$. 

Similarly, the consistency condition for stop is 
	\begin{eqnarray}
	\cos(2\pi M_B R) =\cos(2\pi \alpha \pm 2\pi M_t R) \ ,
	\end{eqnarray}
which leads to mass eigenvalues, 
	\begin{align}
	M_B=\frac{n+\alpha}{R} \pm M_t(H_u) \quad (n:\rm integer). 	 
	\end{align}
This result makes sense because supersymmetric limit, $\alpha=0$, reproduces quark mass spectrum and also because mass splitting of zero mode squarks, $\pm M_t$, is as expected by $A_0=-2\alpha/R$. 

\subsection{Effective Potential}
Once the Higgs-dependent mass spectrum is obtained, we can compute the effective potential, 
	\begin{eqnarray}
	V_t= \frac{2N_c}{2R^4}\sum_{n=-\infty}^\infty \int \frac{d^4 k}{(2\pi)^4} 
	\left\{ \log \frac{k^2 +(n+\omega_{B+})^2}{k^2 +(n+\omega_F)^2}
	 +\log \frac{k^2 +(n+\omega_{B-})^2}{k^2 +(n+\omega_F)^2}
	\right\} \label{eq:Vt}
	\end{eqnarray}
where $\omega_{B\pm}= \alpha \pm M_t$ and $\omega_F=  M_t$. 
Here, we rescale all the dimensionful parameters to be dimensionless by using $R$, for example, $M_t\to M_t/R$.  The numerator of the prefactor, $2N_c$, represents degrees of freedom of a colored complex scalar or a colored Weyl fermion.  

To handle this infinite sum, it is convenient to use $W$ and its derivative defined as  
	\begin{align}
	W(\omega)\equiv \frac{1}{2}\sum_{n=-\infty}^\infty \int \frac{d^4 k}{(2\pi)^4}  \log \frac{k^2 +(n+\omega)^2}{k^2 +n^2},
	\quad
	W'(\omega)=\sum_{n=-\infty}^\infty \int \frac{d^4 k}{(2\pi)^4}  \frac{(n+\omega)}{k^2 +(n+\omega)^2}\ .
	\label{eq:W}
	\end{align}
Then we can rewrite the effective potential of Eq.~\eqref{eq:Vt}, 
	\begin{align}
		V_t &= \frac{2N_c}{R^4} [W(\omega_{B+})+W(\omega_{B-})-2W(\omega_F)]
		\nonumber\\
	&=\frac{2N_c}{R^4} \left[ \int_0^{\omega_{B+}}\!\!\! d\omega\  W'(\omega) 
	 +\int_0^{\omega_{B-}}\!\!\! d\omega\  W'(\omega) 
	 -2\int_0^{\omega_F}\!\!\! d\omega\  W'(\omega) \right].
	\end{align}
$W'$ is computed with a technique well-known in field theory with finite temperature (see Appendix~\ref{app:infinitesum}), 
	\begin{align}
	W'(\omega)&=\frac{-3i}{2(2\pi)^5} [ {\rm Li}_{4}(e^{2\pi i \omega})- {\rm Li}_{4}(e^{-2\pi i \omega})], 
	\end{align}
and hence,
	\begin{align}
	W(\omega)&=\frac{-3}{2(2\pi)^6}  [ {\rm Li}_{5}(e^{2\pi i \omega})+ {\rm Li}_{5}(e^{-2\pi i \omega})]. 
	\end{align}
Polylogarithm, ${\rm Li}_{s}(z)=\sum_{k=1}^\infty \frac{z^k}{k^s}$, implies the KK tower  effect. When $z=1$, it coincides with the Riemann zeta function, $\zeta(s)$.

The effective potential  becomes a simple and finite formula, 
	\begin{align}
	V_t=\frac{-3 N_c}{64\pi^6 R^4} \Big[ {\rm Li}_{5}(e^{2\pi i \omega_{B+}})+ {\rm Li}_{5}(e^{-2\pi i \omega_{B+}}) 
	+ {\rm Li}_{5}(e^{2\pi i \omega_{B-}})+ {\rm Li}_{5}(e^{-2\pi i \omega_{B-}})& \nonumber\\
	-2{\rm Li}_{5}(e^{2\pi i \omega_F})- 2{\rm Li}_{5}(e^{-2\pi i \omega_F})&\Big]\ . \label{eq:Vtfinite}
	\end{align}
 In this calculation, the UV regularization is not needed because the Scherk-Schwarz mechanism, thanks to the non-local supersymmetry breaking, has a noble feature of UV finiteness.  Many literatures \cite{Barbieri:2000vh, ArkaniHamed:2001mi, Ghilencea:2001ug, Delgado:2001ex, Contino:2001gz, Barbieri:2001dm, Masiero:2001im, Delgado:2001xr, Delgado:1998qr, Antoniadis:1998sd, Kim:2001gk} discuss the Scherk-Schwarz mechanism in this context. 
The UV insensitivity is actually observed in intermediated steps of our computation (Appendix~\ref{app:infinitesum}). After the summation of the KK modes, 
there are two pieces:  one has stronger UV divergence  which is independent of $\alpha$ and the other has exponentially suppressed UV dependence.  Since the former is  $\alpha$ independent, the divergence pieces are completely cancelled by combining bosonic and fermionic contributions. 

The similar calculation of the effective potential is found in Ref.~\cite{Cohen:2015gaa} where the Folded Supersymmetry model is used (see also Ref.~\cite{Bagger:2001qi} for a mass spectrum with brane terms). 

 \begin{figure}[t]
 \begin{center}
  \includegraphics[width=0.55\linewidth]{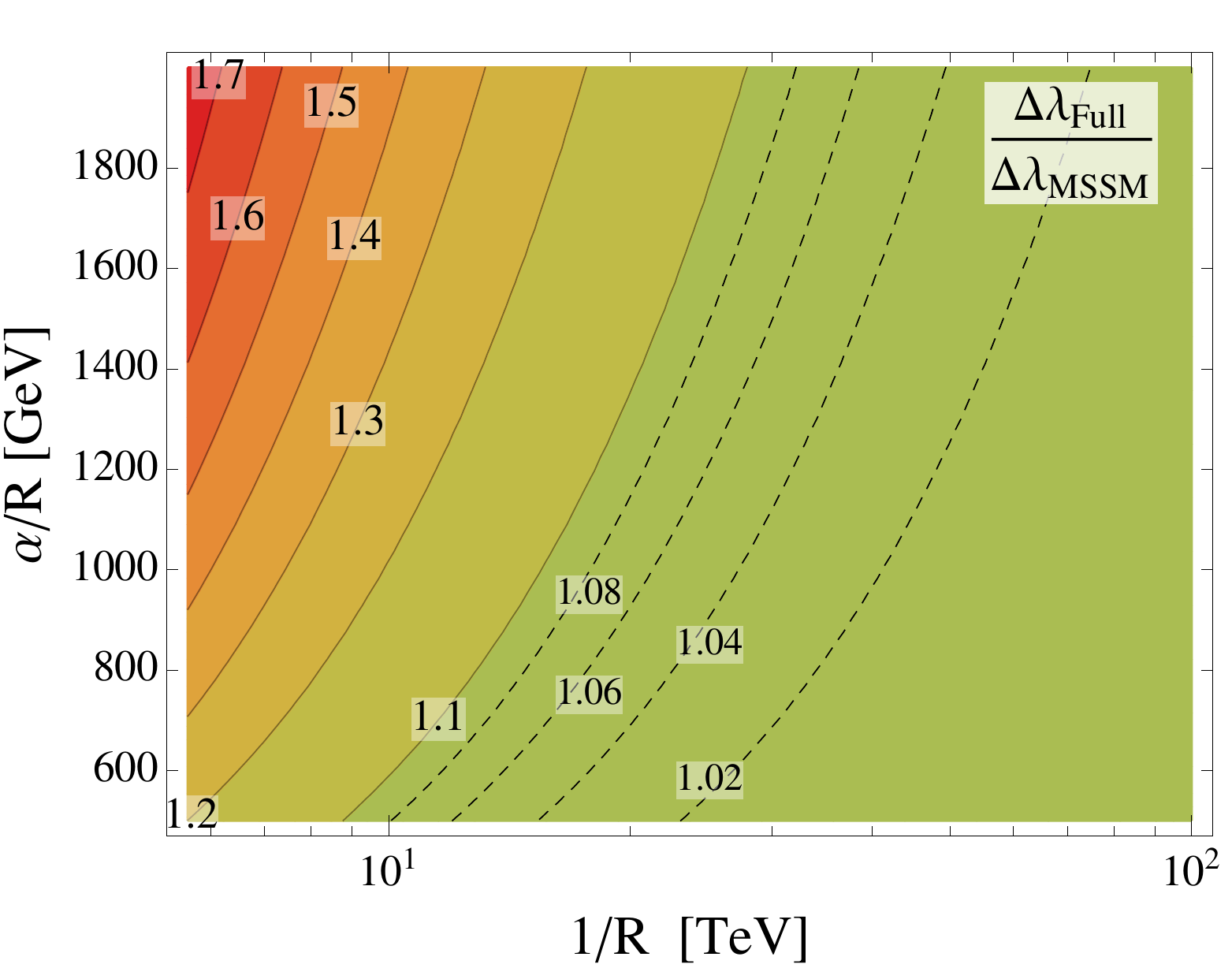}
\vspace{-8pt}
 \end{center}
 \caption{\label{fig:comparison}
 Ratio of 1-loop corrections to $H_u^4$ coupling. $\Delta\lambda_{\rm Full}$ is Higgs quartic coupling generated from all the KK modes, while $\Delta\lambda_{\rm MSSM}$ is that from the zero mode (only the MSSM contribution). }
\end{figure}

\subsection{Enhancement of Higgs Quartic Coupling}
The obtained effective potential is now used to get a new Higgs quartic coupling which includes  the effect of all the KK  modes. 
First, we put $M_t$ back to a dimensionful parameter by extracting $R$ factor and expand the potential with respect to $M_t (\ll \alpha/R)$,
	\begin{align}
	V_t
	=&-\frac{3 N_c}{32\pi^6 R^4}\left({\rm Li}_{5}(e^{2\pi i \alpha }) +{\rm Li}_{5}(e^{-2\pi i \alpha }) -2\zeta(5)	\right)
	\nonumber\\
	&+\frac{3 N_c}{16\pi^4 R^2}\left({\rm Li}_{3}(e^{2\pi i \alpha }) +{\rm Li}_{3}(e^{-2\pi i \alpha }) -2\zeta(3)
	\right)M_t^2 
	\nonumber\\
	&	+\frac{ N_c}{16\pi^2 }\left(
	\frac{25}{6}+\log(1-e^{2\pi i \alpha })(1-e^{-2\pi i \alpha })-2\log(2\pi M_t R)
	\right)M_t^4  
	+\frac{{\cal O}(M_t^6 R^6)}{R^4}.
	\end{align}
The above correction is maximized at $\alpha=1/2$. The term of $M_t^4$ for small $\alpha$ is  
	\begin{align}
	\frac{ N_c}{16\pi^2 }\left(
	\frac{25}{6}+2\log\frac{\alpha/R}{M_t} -\frac{\pi^2 \alpha^2}{3} +{\cal O}(\alpha^4)
	\right) M_t^4. 
	\label{mt4coeff}
	\end{align}
The first two terms give the same result of the Higgs quartic coupling radiatively generated by the MSSM (zero mode) particles. The last term, ${-\pi^2 \alpha^2}/{3}$,  is  from the KK tower but it is negative, which decreases the Higgs mass. 
However, an important effect of the KK tower comes from $M_t^2$ term since $M_t$ has higher power of $H_u$ as shown in Eq.~\eqref{topmass}. The potential is expanded with respect to $H_u$, 
	\begin{align}
	V_t	=\ &-\frac{3 N_c}{32\pi^6 R^4}\left({\rm Li}_{5}(e^{2\pi i \alpha }) +{\rm Li}_{5}(e^{-2\pi i \alpha }) -2\zeta(5)	\right)
	\nonumber\\
	\	&+
	\frac{3y_t^2 N_c}{16\pi^4 R^2}\left({\rm Li}_{3}(e^{2\pi i \alpha }) +{\rm Li}_{3}(e^{-2\pi i \alpha }) -2\zeta(3)
	\right) H_u^2 
	\nonumber\\
	&	+\frac{y_t^4 N_c}{16\pi^2 }\bigg(
	\frac{25}{6}+\log(1-e^{2\pi i \alpha })(1-e^{-2\pi i \alpha })-2\log(2\pi M_t R)
	\nonumber\\
	&\hspace{3.7cm}  -2{\rm Li}_{3}(e^{2\pi i \alpha })-2{\rm Li}_{3}(e^{-2\pi i \alpha })+4\zeta(3)
	\bigg)H_u^4  
	\ \ +\frac{{\cal O}(H_u^6 R^6)}{R^4}
	\label{Huexpand}	.
	\end{align}
The last line is from $M_t^2$ term, and it gives a large and positive contribution, 
	\begin{eqnarray}
	-2{\rm Li}_{3}(e^{2\pi i \alpha })-2{\rm Li}_{3}(e^{-2\pi i \alpha })+4\zeta(3)
	=\left(12 -8\log (2\pi \alpha) \right)\pi^2\alpha^2  +{\cal O}(\alpha^4). 
	\end{eqnarray}
For instance, combining with the last term  in Eq.~\eqref{mt4coeff}, $-\pi^2\alpha^2/3$, a total contribution to the Higgs quartic coupling from the KK tower is proportional to $(-1/3+12 -8\log (2\pi \alpha))\pi^2 \alpha^2 =3.3\times 10^{-2}~\text{--}~3.9$ for parameter of our interest, $\alpha=10^{-2}~\text{--}~0.2$, which can be compared to the MSSM contribution $\approx 25/6$. 
To see the impact of this new effect, Fig.~\ref{fig:comparison} shows a comparison between $\Delta\lambda_{\rm Full}$ defined as the Higgs quartic coupling in Eq.~\eqref{Huexpand} and 
the Higgs quartic coupling in the MSSM, $\Delta\lambda_{\rm MSSM}$,  defined by
	\begin{align}
	\Delta\lambda_{\rm MSSM}H_u^4 \equiv 
		\frac{y_t^4 N_c}{16\pi^2 }\left(
	\frac{25}{6}+2\log\frac{\alpha/R}{M_t} 	\right) H_u^4 
	,
	\label{HuexpandMSSM}
	\end{align}
where $M_t \ll \alpha/R$ is assumed.
It is shown in Fig.~\ref{fig:comparison} that 
even for a 10 TeV scale of the extra dimension, the effect of the KK modes enhances $10\sim50\%$ radiative correction to the quartic coupling.  This is surprising because, if the MSSM has mass scale of $\alpha/R\sim \rm TeV$, modification from the KK modes is naively expected to be $(\alpha/R)^2/R^{-2}\sim 1\%$ rather than ${\cal O}(10\%)$. 
Therefore this new contribution will give a big change to the calculation of the lightest Higgs mass. 

Note that terms with higher power of $H_u$, such as $H_u^6$, are not important for the Higgs mass when we consider parameter space of $v\ll R^{-1}$ hence $H_u R$ expansion is valid. We explicitly show this in Appendix.~\ref{app:higherterm}.

\section{Higgs Mass}\label{sec:Higgsmass}
We found the  Higgs quartic coupling enhanced by the full KK tower. Then we need electroweak symmetry breaking parameters, such as $\tan\beta\equiv\langle H_u\rangle/\langle H_d\rangle$, to evaluate the lightest Higgs mass, $m_h$. In the following Sec.~\ref{sec:EWSB}, we obtain $\mu$ and $\tan\beta$ by solving electroweak symmetry breaking conditions. 
In Sec.~\ref{sec:HOC}, we improve 1-loop calculation from the top sector by RGE to include next-to-leading logarithm order and also take into account the leading correction of electroweak gauge couplings. Finally, in Sec.~\ref{sec:enhancedHiggsmass}, we present results of the Higgs mass and fine-tuning. 

\subsection{Electroweak Symmetry Breaking}\label{sec:EWSB}
\begin{figure*}[h]
 \begin{center}
  \includegraphics[width=0.55\linewidth]{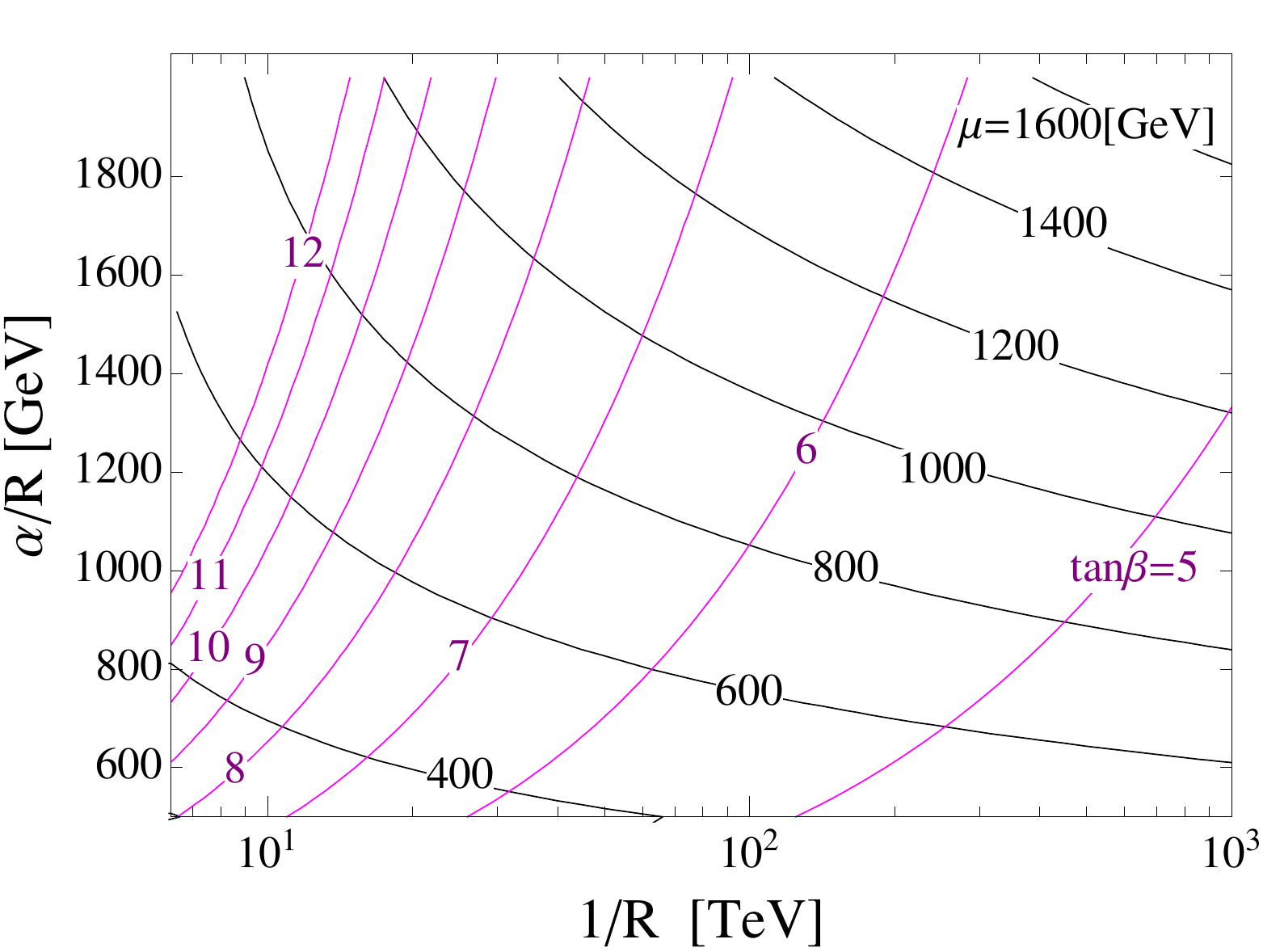}
\vspace{-8pt}
 \end{center}
 \caption{$\mu$ and  $\tan\beta$ are determined by requiring the successful electroweak symmetry breaking.  
 Black lines show $\mu$ in GeV and purple lines show $\tan\beta$. 
 \label{mutanb}}
\end{figure*}

Despite the constrained structure of the model, successful electroweak symmetry breaking can be achieved.  Since the scale of the extra dimension is well above the supersymmetry breaking scale, we match the theory onto the MSSM with a matching (renomalization) scale, 
	\begin{eqnarray}
	Q_{\rm RG}=\frac{1}{2\pi R}\ .
	\end{eqnarray}
Higgs soft terms are absent at tree level but  are generated radiatively. 
We  calculate those corrections including all the KK modes which are finite thanks to the Scherk-Schwarz mechanism, and subtract the MSSM parts which are regularized with the $\overline{\rm DR}$ scheme from them.  As a result, threshold corrections to the Higgs soft terms at the matching scale are 
\begin{eqnarray}
  m_{H_u}^2 
  &=& \left( -\frac{3 y_t^2}{\pi^2} 
    + \frac{3 (g_2^2 + g_1^2/5)}{8\pi^2} \right) 
    \left( \frac{\alpha}{R} \right)^2,
\\[5pt]
  m_{H_d}^2 
  &=& \frac{3 (g_2^2 + g_1^2/5)}{8\pi^2} 
    \left( \frac{\alpha}{R} \right)^2,
\\[5pt]
  b 
  &=& \left( \frac{3 y_t^2}{4\pi^2} 
    - \frac{3 (g_2^2 + g_1^2/5)}{16\pi^2} \right) 
    \mu\frac{\alpha}{R}.
\label{eq:corr-2}
\end{eqnarray}
The detail of these results is given in Appendix~\ref{app:threshold}. 
We solve electroweak symmetry breaking conditions using  {\tt SOFTSUSY~3.4} \cite{Allanach:2001kg}. Among three free parameters of the model, one of them is determined by the Higgs VEV. In Fig.~\ref{mutanb}, $\mu$ as well as $\tan\beta$ are shown in $\alpha/R$ and $1/R$ parameter space.  
 
We find the Higgsino mass scale, $\mu$, has to be close to gaugino, squark, and slepton mass scale, $\alpha/R$, which leads to a compressed spectrum which  ameliorates  LHC bounds. The spectrum is more compressed as $R^{-1}$ gets large because $\mu$ grows to  compensate $-|m_{H_u}^2|$ which becomes larger due to a long  RGE running. 
Since $\tan\beta$  is found to be as low as $\tan\beta\lesssim 10$ in the parameter space, the tree level Higgs mass, $|m_Z \cos(2\beta)|$,  is rather low and hence a large radiative correction is needed to realize the observed value $m_h\approx125$~GeV. 

\subsection{Higher Order Corrections}\label{sec:HOC}
It is known that the corrections at ${\cal O}(y_t^4 g_s^2, y_t^6)$ are significant and this is mainly due to the scale of the top mass. Here we adopt a RG-improved method \cite{Haber:1996fp} (see also Refs.~\cite{Haber:1993an,Carena:1995bx, Carena:2000yi}) based on our 1-loop calculation. The top mass (top Yukawa and $H_u$ VEV) runs to an intermediate scale of top quark and stop by 
	\begin{eqnarray}
	y_t\to y_t(M_t) \left(1+\beta_t\log\frac{Q}{M_t}	\right), \quad\ 
	v_u\to v_u(M_t)\left(1+\gamma_{v_u} \log\frac{Q}{M_t}\right)\ ,
	\label{topYukawaRun}
	\end{eqnarray}
where 
	\begin{align}
	Q=c_{t}\sqrt{M_t(M_t^2+\alpha^2/R^2)^{1/2}}\ .
	\end{align}
The beta function and anomalous dimension are those for the two-Higgs Doublet Model, 
	\begin{eqnarray}
	\beta_t=\frac{9y_t^2}{32\pi^2}-\frac{g_s^2}{2\pi^2}, \quad \gamma_{v_u}=-\frac{3y_t^2}{16\pi^2} . 
	\end{eqnarray}
We choose $c_{t}=2.1$ so that this RG-improved calculation for the MSSM Higgs mass  reproduces the 2-loop result calculated by {\tt FeynHiggs} \cite{Heinemeyer:1998yj} to a good accuracy. This top mass is used for the 
formulae for the Higgs quartic coupling in Eqs.~(\ref{Huexpand}, \ref{HuexpandMSSM}). 
Regarding the KK mode contribution, say $\Delta \lambda_{\rm KK} H_u^4 \equiv (\Delta \lambda_{\rm Full}-\Delta \lambda_{\rm MSSM})H_u^4$, a more appropriate treatment  is that $\Delta \lambda_{\rm KK} H_u^4$ is given at the KK scale of $1/R$ and runs down to the scale $\alpha/R$. However, since such an effect  is subdominant and requires to consider mixing with other operators, it is beyond the scope of this paper.  

In order to  further improve our computation to the Higgs mass, we include ${\cal O}(y_t^2 g_1^2, y_t^2 g_2^2)$ correction. Here, we focus on the MSSM contribution and the corresponding correction to the Higgs potential is 
	\begin{eqnarray}
	V_{y_t^2 g^2}
	=-\frac{g_1^2+g_2^2}{4}\frac{3y_t^2}{8\pi^2} \log \frac{m_{\tilde{t}}^2 +M_t^2}{M_t^2} 
	\left( |H_u|^4 -\frac{1}{2} |H_u|^2|H_d|^2 \right) \ ,
	\end{eqnarray}
where we neglect the mixing between $\tilde{t}_R$ and $\tilde{t}_L$. The top Yukawa here is also given by Eq.~\eqref{topYukawaRun}. 

 \begin{figure*}[t]
 \begin{center}
  \includegraphics[width=0.55\linewidth]{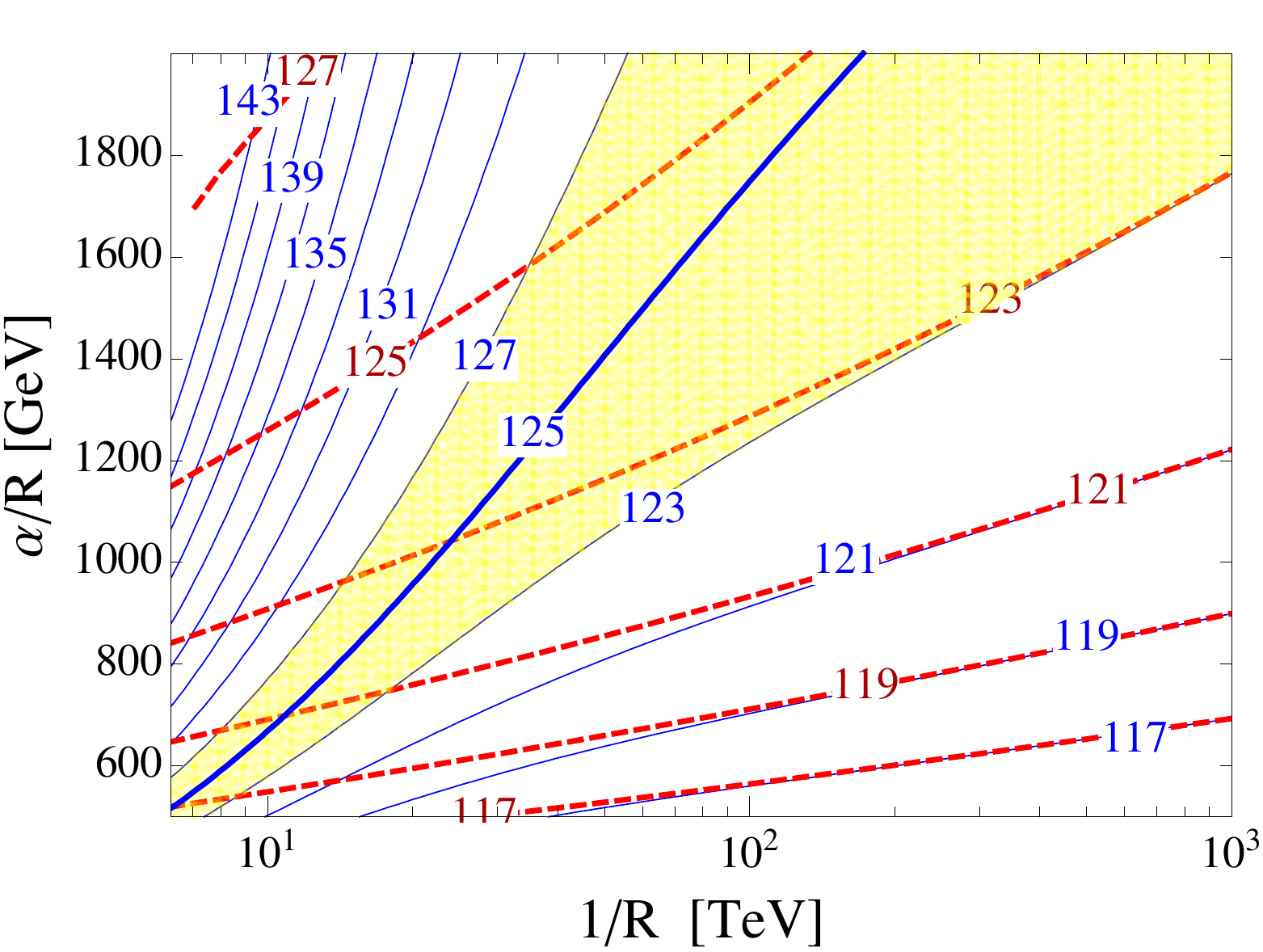}
\vspace{-8pt}
 \end{center}
 \caption{ The lightest Higgs mass in an unit of GeV. Each blue solid line is Higgs mass calculated with the full KK tower, and each red dashed line corresponds to Higgs mass based on the MSSM calculation.  The MSSM calculation always underestimates the Higgs mass. 
The yellow band is a region of $m_h=125\pm 2~{\rm GeV}$
 \label{fig:Higgsmass}
}
\end{figure*}

\subsection{Enhanced Higgs Mass}\label{sec:enhancedHiggsmass}
Based on the RG-improved method and electroweak parameters, we calculate the Higgs mass. 
In Fig.~\ref{fig:Higgsmass}, we show the Higgs mass calculated at the MSSM level and that calculated with the full KK tower. 
Even for heavy KK modes of $R^{-1}\sim 10~\rm TeV$, a line of $m_h=125~\rm GeV$ based on the MSSM is underestimated because once the full KK tower is included, the Higgs mass is significantly enhanced and then the  line of $m_h=125~\rm GeV$ based on the MSSM actually corresponds to  $m_h=$130~--~140~GeV. 
The true line of $m_h=125~\rm GeV$ is realized in a lower supersymmetry breaking scale. 
The line becomes  a band if we consider an uncertainty of our prediction, and we take conservatively $2~\rm GeV$ as an uncertainty of the Higgs mass which corresponds to a band in Fig.~\ref{fig:Higgsmass}.
   As in Fig.~\ref{fig:comparison} of the Higgs quartic coupling, the effect of KK modes eventually disappears when $R^{-1}$ goes beyond 100~TeV ($\alpha\lesssim 0.01$). 

 Improvement of Higgs mass calculation is very important in the Compact Supersymmetry model because it has only three free parameters, $R$, $\alpha/R$, and $\mu$, one of which is determined by the Higgs VEV, and furthermore the observed Higgs mass constraints one more parameter leading to just a line (a band with an error) in the parameter space. It is shown that the MSSM calculation points to wrong region in the parameter space. 
  The improved calculation tells that supersymmetry breaking scale that explains the Higgs mass is lowered, which motivates TeV supersymmetry signature with a compressed spectrum. Also,  testability of the model at the LHC increases.

\begin{figure*}[t]
 \begin{center}
  \includegraphics[width=0.55\linewidth]{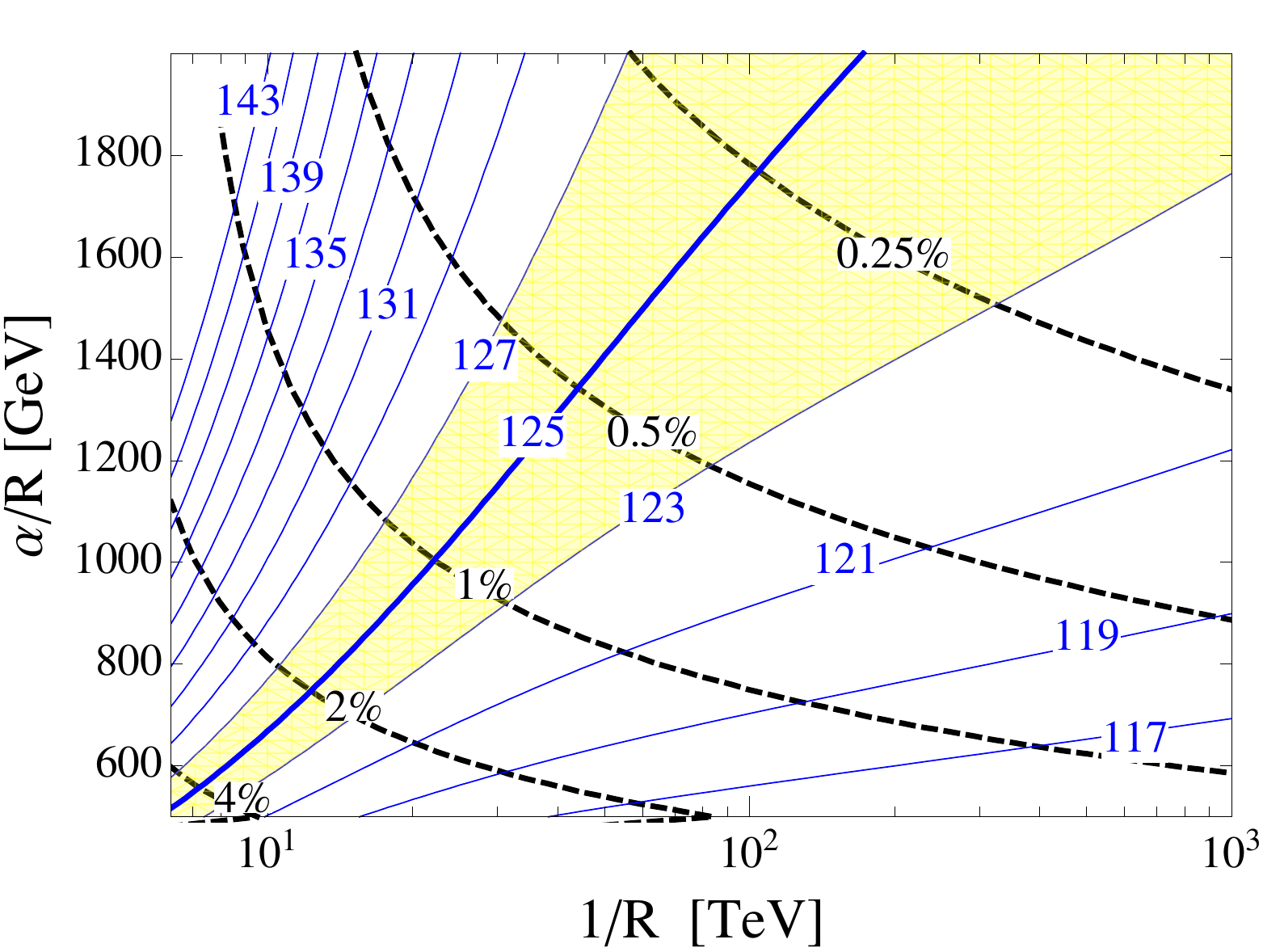}
\vspace{-8pt}
 \end{center}
 \caption{ 
 Fine-tuning $\Delta_\mu^{-1}\approx m_h^2/(4\mu^2)$ is plotted as black dashed lines. Blue solid lines correspond to  the lightest Higgs mass in a unit of GeV calculated with the full KK tower, and the yellow band is a region of $m_h=125\pm 2~{\rm GeV}$. 
 \label{fig:tuning}
}
\end{figure*}

Fine-tuning is also investigated. For the successful electroweak symmetry breaking, $\mu$ has to be almost as big as the supersymmetry breaking scale, $\alpha/R$, which leads to a tree level tuning of the weak scale due to $\mu\gg v, m_Z$. 
Since $\mu$ is the dominant source of fine-tuning to realize the correct weak scale as pointed out in Ref.~\cite{Murayama:2012jh},  we adopt a simple fine-tuning measure, $\Delta_\mu^{-1}$, varying only $\mu$, 
	\begin{align}
	\Delta_\mu^{-1}=\left|\frac{\partial\log v^2}{\partial\log \mu}\right|^{-1} \approx \frac{m_h^2}{4\mu^2}\ . 
	\end{align}
The last approximation is valid when heavy Higgs states are decoupled \cite{Kitano:2006gv}. In this case, the Higgs sector corresponds to SM-like one-Higgs doublet model, and  we can derive 
	\begin{align}
	\frac{\partial v^2}{\partial \mu^2} =\frac{v^2}{m_h^2}\frac{\partial m_h^2}{\partial \mu^2}=\frac{-2v^2}{m_h^2}\ .
	\end{align}
Using this approximation, in Fig.~\ref{fig:tuning},  we show   fine-tuning of the Compact Supersymmetry model is percent level.

\section{Experimental Bounds}\label{sec:Expbound}
In this section, we discuss how experimental results bound on the parameter space. The first constraint we consider is from the electroweak precision test. Since this model has brane-localized Higgs, the Higgs VEV mixes zero mode and non-zero modes of electroweak gauge bosons at tree level. Refs.~\cite{Delgado:1999sv, Delgado:2001si} study such bounds and lead to a limit on a size of the extra dimension, $R^{-1}\gtrsim 5\rm~TeV$. It does not  constrain interesting parameter space which can explain the Higgs mass and the current LHC results.

\begin{figure*}[t]
 \begin{center}
  \includegraphics[width=0.55\linewidth]{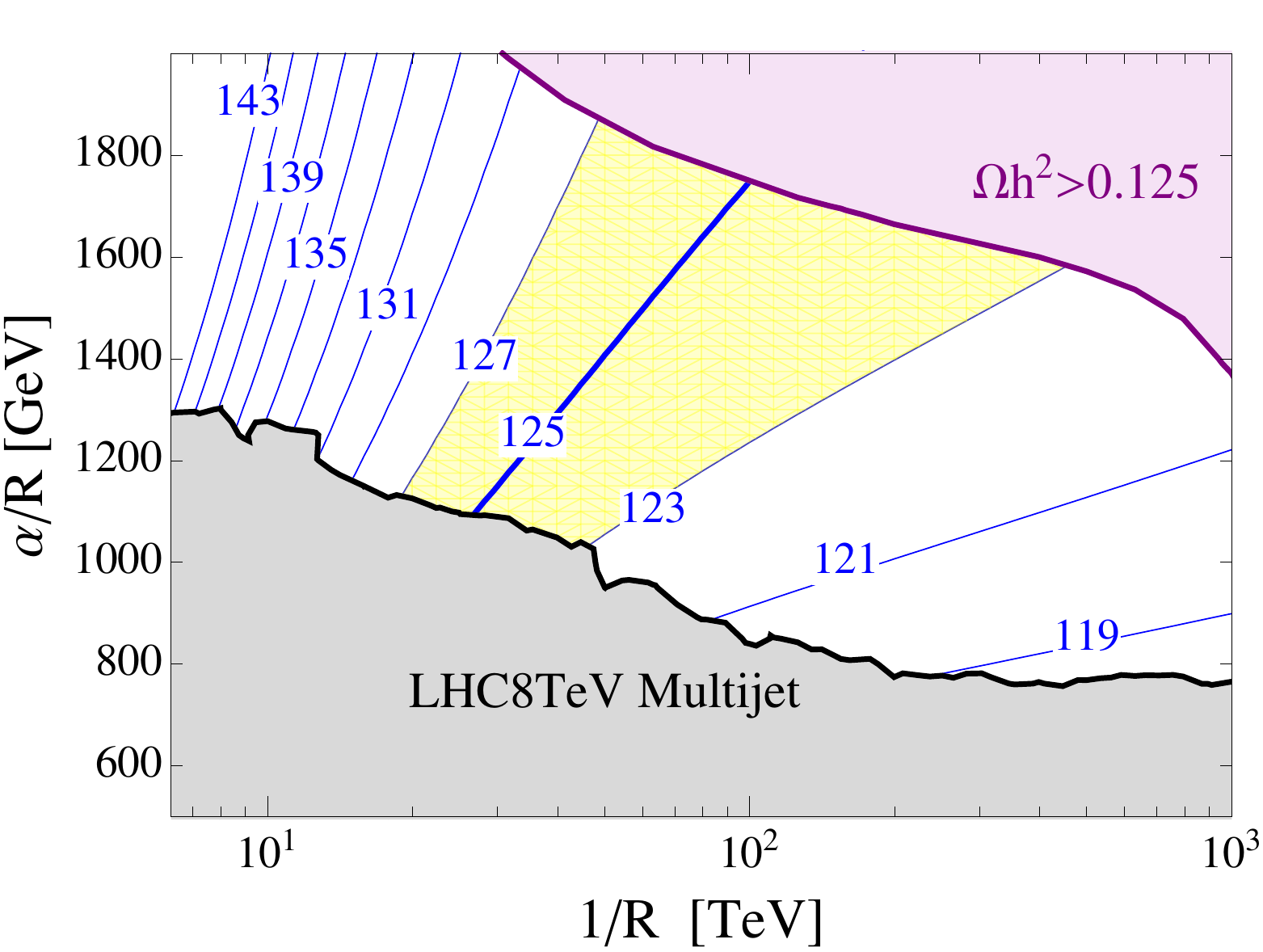}
\vspace{-8pt}
 \end{center}
 \caption{ The upper (purple) shaded  region is excluded by the thermal relic abundance of the LSP larger than the observed dark matter relic abundance, $\Omega h^2>0.125$.  The lower (gray) shaded region is excluded by one of the ATLAS results which are based on multijet+$\slashed{E}_T$.    
 Blue solid lines correspond to  the lightest Higgs mass in a unit of GeV calculated with all the KK tower, and the yellow band is a region of $m_h=125\pm 2~{\rm GeV}$. 
 \label{fig:bound}
}
\end{figure*}

Secondly, the ATLAS and CMS experiments search for supersymmetric particles and derive bounds in many channels. Here we study a representative analysis of multijet+$\slashed{E}_T$. Since the Compact Supersymmetry model  has a compressed spectrum whose mass difference between the LSP and gluino/squark is typically 300 GeV, the bound is weaker  than those for the CMSSM and simplified models ($m_{\tilde{g}}\approx m_{\tilde{q}}$)~\cite{Aad:2014wea}. For the spectrum calculation, as discussed in Sec.~\ref{sec:EWSB}, we match the theory onto the MSSM at a scale of $1/(2\pi R)$ and consider RGE running down to the supersymmetry breaking scale, $\alpha/R$. 
We generate signal events by {\tt PYTHIA~6.4}~\cite{Sjostrand:2006za},  and  use {\tt PGS~4}~\cite{PGS} for the detector simulation and {\tt NLL-fast}~\cite{NLLFAST, Beenakker:1996ch, Kulesza:2008jb, Kulesza:2009kq, Beenakker:2009ha, Beenakker:2011fu} for 
estimation of the production cross section including next-to-leading order QCD corrections and the resummation at next-to-leading-logarithmic accuracy. 
We compare the obtained event numbers with ATLAS searches using multijet +$\slashed{E}_T$  without lepton with ${\cal L} = 20.3~{\rm fb}^{-1}$ at $\sqrt{s} = 8~{\rm TeV}$~\cite{Aad:2014wea}, and the result is shown as a lower shaded region in Fig.~\ref{fig:bound}.
We find the exclusion bound is extended up to $m_{\tilde{g}}\simeq 1\rm~TeV$, and for a region at $R^{-1}\sim 10\rm~TeV$ the bound is stronger as $m_{\tilde{g}}\gtrsim 1.3\rm~TeV$ because the spectrum is less compressed in this region. 
In contrast, the CMSSM and simplified model are more constrained as the bound is $m_{\tilde{g}}\gtrsim 1.7\rm~TeV$. 

Finally, the LSP can be a dark matter candidate. As long as the LSP is stable,  its relic abundance should be lower than the observed dark matter relic abundance. We calculate thermal relic abundance of the LSP using {\tt MicroOMEGAs}~\cite{Belanger:2006is, Belanger:2008sj} , and put a 95\%~C.L. upper bound on the relic abundance, $\Omega h^2<0.125$, obtained by Planck collaboration \cite{Ade:2013zuv}.  The LSP in this model is dominated by the Higgsino component because of  $\mu<\alpha/R$, and then the thermal relic abundance of the LSP is smaller than the observed abundance unless it is heavy to  decouple from thermal bath earlier.  
 The excluded region is shown as an upper shaded region in Fig.~\ref{fig:bound}. It is very interesting because this tells that the model can be tested, that is, along the Higgs mass band there is upper bound in the TeV range by the dark matter relic abundance and the LHC bound from the bottom will be improved at upcoming LHC run at $\sqrt{s}=$13 and 14~TeV.

Regarding the future search for models with a compressed spectrum such as the Compact Supersymmetry model, since the signal is weaker, it is important to improve the sensitivity. One possibility is to utilize $M_{T2}$~\cite{Lester:1999tx} which can systematically separate signal and background, and  its validity is  demonstrated  in Ref.~\cite{Murayama:2011hj} that $M_{T2}$  significantly improves discovery potential of the Minimal Universal Extra Dimension \cite{Appelquist:2000nn} which has a compressed spectrum.  Also,  other useful techniques~\cite{Alwall:2008va, Rolbiecki:2012gn, Dreiner:2012gx, Dreiner:2012sh, Bhattacherjee:2013wna, Mukhopadhyay:2014dsa, Han:2015lha} are developed to improve the sensitivity to models with compressed spectra at the LHC.  Using these techniques and all channels of experiments, we believe the whole parameter space of the Compact Supersymmetry model compatible with the observed Higgs mass and the dark matter relic abundance ($\alpha/R \lesssim 1.8$~TeV) is explicitly testable at the LHC.

\section{Conclusions}\label{sec:conclude}
We studied an impact of the KK tower to the lightest Higgs mass in the Compact Supersymmetry model. We computed the effective potential of all the KK modes (with Higgs dependent mass eigenvalues), and find the enhancement of the Higgs quartic coupling is unexpectedly large. The effect of the KK modes enhances the radiative contribution to the Higgs quartic coupling by from 10 to 50~\% even for heavy KK modes of $\cal O$(10) TeV, and the effect remains non-negligible until $\cal O$(100) TeV.  
This is mainly because the top quark wave function is pushed out from the brane, which makes the top mass depend on higher powers in the Higgs field.
Correspondingly, the Higgs mass is raised by from 5 to 15 GeV, and hence the Higgs mass of 125 GeV is realized in a lower supersymmetry breaking scale of $\alpha/R$. 
 The better knowledge of the Higgs mass together with the Higgs VEV essentially leaves only one free parameter of the model. 
Furthermore the parameter space is bounded, with respect to $\alpha/R$, from the bottom by the LHC searches at $\alpha/R\simeq1$~TeV and  from the top by the the dark matter relic abundance at $\alpha/R\simeq1.8$~TeV. 
 Although the compressed spectrum weakens LHC bounds, since the LHC run~II can investigate higher mass scale by the higher energy,  the whole parameter space of the model will be explicitly tested.

\section*{Acknowledgement}
We thank Lorenzo di Pietro, Hou Keong Lou, Xiaochuan Lu, Yasunori Nomura, and Ryosuke Sato for useful discussions. 
This work was supported by JSPS KAKENHI Grant-in-Aid for Scientific Research (B) (No.~15H03669 [RK]) and (C) (No.~26400241 [HM]), Grant-in-Aid for JSPS Fellows (No.~14J00179 [KT]), MEXT KAKENHI Grant-in-Aid for Scientific Research on Innovative Areas (No.~25105011 [RK], No.~15H05887 [HM]),  and by WPI, MEXT, Japan. 
HM also was supported in part by the U.S. DOE under Contract DE-AC03-76SF00098, in part by the NSF under grant PHY-1316783.

\appendix
\section{Mass Spectrum}\label{app:spectrum}
\subsection{Equations of Motion}\label{app:EOM}
In order to obtain Higgs-dependent mass eigenvalues, we solve 5D equations of motion and find wave functions.  
Equations of motion for squark are
	\begin{eqnarray}
	&&(-\partial^2+\partial^2_5)\tilde{Q} -y_{U5}^2 H_u^2 \delta(y)\delta(0)\tilde{Q}
	+y_{U5} H_u \delta(y)\partial_5 \tilde{U}^{c}=0  \label{EOM1}, 
	\\
	&&(-\partial^2+\partial^2_5)\tilde{U}^{c}-y_{U5} H_u \partial_5[\delta(y) \tilde{Q} ]=0 \label{EOM2}, 
	\\
	&&(-\partial^2+\partial^2_5)\tilde{U}^* -y_{U5}^2 H_u^2 \delta(y)\delta(0)\tilde{U}^*
	+y_{U5} H_u \delta(y)\partial_5 \tilde{Q}^{c*}=0  \label{EOM3}, 
	\\
	&&(-\partial^2+\partial^2_5)\tilde{Q}^{c*}-y_{U5} H_u \partial_5[\delta(y) \tilde{U}^* ]=0 \label{EOM4}. 
	\end{eqnarray}
Since we are only interested in top quark, $\tilde{Q}^{(c)}$ represents 5D top squark field in the $SU(2)_L$ doublet and  $\tilde{U}^{(c)}$ also represents 5D top squark field,  and therefore Yukawa coupling here corresponds to top Yukawa, $y_{U5}\to y_{t5}$.  
All the parameters are taken to be real for simplicity. In the following, we perform 4D Fourier transformation and consider on-shell, that is, $-\partial^2\to p^2=M^2$. 

Equations of motion for quark are
	\begin{eqnarray}
	(i\slashed{\partial} +\gamma_5 \partial_5)\Psi_Q +\delta(y)y_{U5}P_R\Psi_U^C H_u=0,
	\label{EOM5}
	\\
	(i\slashed{\partial} +\gamma_5 \partial_5)\Psi_U +\delta(y)y_{U5}P_R\Psi_Q^C H_u=0,
	\label{EOM6}
	\end{eqnarray}
where the superscript $C$ denotes charge conjugation of the fermion. 
As in the squark case, we focus on top quark.

\subsection{Solution for Quark}\label{app:quark}
We can separate quark wave functions to 4D parts and extra dimensional parts, 
	\begin{eqnarray}
	\Psi_Q(x,y)=\left(\begin{array}{c}{Q}(x,y) \\  \overline{Q}^c(x,y)\end{array}\right)
	= \psi_{Q-}(x)f_{Q-}(y)+\psi_{Q+}(x)f_{Q+}(y), 
	\\
	\Psi_U(x,y)=\left(\begin{array}{c}{U}(x,y) \\  \overline{U}^c(x,y)\end{array}\right)=
	 \psi_{U-}(x)f_{U-}(y)+\psi_{U+}(x)f_{U+}(y). 
	\end{eqnarray}
 $\psi_{\pm}$ is four component spinor which has a chirality, $\gamma_5 \psi_{\pm}=\pm \psi_\pm$. 
Under the reflection, ${\cal Z}:y\to -y$, $f_{Q-}(y)$  and $f_{Q+}(y)$ should have the same transformations of $Q(x,y)$ and $\overline{Q}^c(x,y)$, respectively, as 
	\begin{eqnarray}
	f_{Q-}(-y)=+f_{Q-}(y), \quad f_{Q+}(-y)=-f_{Q+}(y) .
	\end{eqnarray}
This condition is same for $U$ quark fields. 

Now we investigate the bulk Lagrangian omitting arguments for simplicity, 
	\begin{align}
	\overline{\Psi}_Q(i\slashed{\partial} +\gamma_5 \partial_5)\Psi_Q  \nonumber
	=&
	\left(\bar\psi_{Q-}f_{Q-}^*+\bar\psi_{Q+}f_{Q+}^*\right)
	(i\slashed{\partial} +\gamma_5 \partial_5)\left(\psi_{Q-}f_{Q-}+\psi_{Q+}f_{Q+}\right)
	\\
	=&(\bar\psi_{Q-}i\slashed{\partial} \psi_{Q-})(f_{Q-}^*f_{Q-}) 
	+	(\bar\psi_{Q+}i\slashed{\partial} \psi_{Q+})(f_{Q+}^*f_{Q+}) \nonumber
	\\&+(\bar\psi_{Q-}\gamma_5 \psi_{Q+})(f_{Q-}^* \partial_5f_{Q+}) 
	+	(\bar\psi_{Q+}\gamma_5 \psi_{Q-})(f_{Q+}^* \partial_5f_{Q-}) \ .
	\end{align}
EOMs in the bulk lead to 
	\begin{eqnarray}
	i\slashed{\partial} \psi_{Q-}(f_{Q-}^*f_{Q-})&=&-\psi_{Q+}(f_{Q-}^* \partial_5f_{Q+}),
	\\
	i\slashed{\partial} \psi_{Q+}(f_{Q+}^*f_{Q+})&=&\psi_{Q-}(f_{Q+}^* \partial_5f_{Q-}) .
	\end{eqnarray}
The last equality is obtained by integration by parts. Separation of variables is used, 
	\begin{align}
	\frac{i\slashed{\partial} \psi_{Q-}(x)}{\psi_{Q+}(x)}&=M=-\frac{ \partial_5f_{Q+}(y)}{f_{Q-}(y)},
	\label{bunri1}\\
	\frac{i\slashed{\partial} \psi_{Q+}(x)}{\psi_{Q-}(x)}&=M'=\frac{ \partial_5f_{Q-}(y)}{f_{Q+}(y)}.
	\label{bunri2}
	\end{align}
$M$ and $M'$ constant and interpreted as 4D quark mass, and  $M=M'$ is necessary because $\psi_{Q+}$ and $\psi_{Q-}$ behave as a single Dirac fermion.  
Then, a differential equation for $f_{Q-}(y)$ is 
	\begin{eqnarray}
	\partial_5^2 f_{Q-}(y) =\partial_5(Mf_{Q+}(y))=-M^2f_{Q-}(y), 
	\end{eqnarray}
and a general solution of $f_{Q-}$ in the bulk is
	\begin{eqnarray}
	f_{Q-}(y)=A\cos(My)+B{\rm sign}(y) \sin(M y) . 
	\end{eqnarray}
A solution of $f_{Q+}$ is  given by $f_{Q+}(y)=\partial_5f_{Q-}(y)/M$, 
	\begin{eqnarray}
	f_{Q+}(y)=-A\sin(My)+B{\rm sign}(y) \cos(M y) . 
	\end{eqnarray}
For $U$ quark, we have similar conditions, 
\footnote{The reason that the constant $M$ here is common with one for $Q$ is because we later obtain conditions that $\psi_{Q}$ and $\psi_{U}^c$ are related in Eqs.~(\ref{quark:constraint3}, \ref{quark:constraint4}).
In other words, they are mixed by Higgs VEV. }
	\begin{eqnarray}
	M=\frac{i\slashed{\partial} \psi_{U-}(x)}{\psi_{U+}(x)}=-\frac{ \partial_5f_{U+}(y)}{f_{U-}(y)}
	=\frac{i\slashed{\partial} \psi_{U+}(x)}{\psi_{U-}(x)}=\frac{ \partial_5f_{U-}(y)}{f_{U+}(y)}	\label{bunri3}
	\end{eqnarray}
and here are solutions, 
	\begin{eqnarray}
	f_{U-}(y)&=&C\cos(My)+D{\rm sign}(y) \sin(M y) , \\
	f_{U+}(y)&=& -C\sin(My)+D{\rm sign}(y) \cos(M y) . 
	\end{eqnarray}

For the case of quark, we can define another reflection, ${\cal Z}': y+\pi R \to -y+\pi R$, that is a product of $\cal T$ and $\cal Z$ (${\cal Z '}={\cal Z T}$). For fermions of $\Phi^c$ which are odd under ${\cal Z}$, they are also odd under ${\cal Z}'$ because of ${\cal T}=1$. In particular, $Q^c(\pi R)=U^c(\pi R)=0$ gives
	\begin{eqnarray}
	A\sin(M\pi R)=B\cos(M\pi R),  \label{Z'Qc}\\
	C\sin(M\pi R)=D\cos(M\pi R). \label{Z'Uc}
	\end{eqnarray}

We determine the coefficients by integrating EOM of Eq.~\eqref{EOM5} around $y=0$ with an infinitesimal interval of $\epsilon$,
	\begin{eqnarray}
	0=\int_{-\epsilon}^\epsilon \!\!\!dy \left\{
	(i\slashed{\partial} +\gamma_5 \partial_5)\Psi_Q(y) +\delta(y)y_{U5}P_R\Psi_U^C(y) H_u
	\right\} \label{bc:quark1} \ .
	\end{eqnarray}
The 4D kinetic term does not give a constraint because  it automatically vanishes,
	\begin{eqnarray}
	\int_{-\epsilon}^\epsilon \!\!\!dy \ i\slashed{\partial}\Psi_Q(y)= 
	\int_{-\epsilon}^\epsilon \!\!\!dy\left\{ M\psi_{Q+} f_{Q-}(y)+ M\psi_{Q-} f_{Q+}(y)\right\}=0 \ .
	\end{eqnarray}
We focus on the right-handed component of Eq.~(\ref{bc:quark1}), 
	\begin{align}
		0&=\int_{-\epsilon}^\epsilon \!\!\!dy \left\{
	\psi_{Q+}\partial_5 f_{Q+}(y) +\delta(y)y_{U5} \psi_{U-}^C f_{U-}^*(y) H_u  \nonumber
	\right\}\\
	&=\psi_{Q+}f_{Q+}(\epsilon)-\psi_{Q+}f_{Q+}(-\epsilon)+y_{U5} \psi_{U-}^C f_{U-}^*(0) H_u  \nonumber\\
	&=2\psi_{Q+} B + y_{U5} H_u \psi_{U-}^C  C^*.  \label{quark:constraint3}
	\end{align}
The other fermion EOM of Eq.~\eqref{EOM6} gives a similar condition, 
	\begin{eqnarray}
	2\psi_{U+} D + y_{U5} H_u \psi_{Q-}^C  A^*=0.  \label{quark:constraint4}
	\end{eqnarray}
This is modified using Eq.~(\ref{bunri3}), 
	\begin{eqnarray}
	2i\slashed{\partial}\psi_{U+} D + y_{U5} H_u i\slashed{\partial}\psi_{Q-}^C  A^*
	=
	2M\psi_{U-} D + y_{U5} H_u M\psi_{Q+}^C  A^*=0  \ . \label{quark:constraint5}
	\end{eqnarray}
Finally, Eqs.~(\ref{Z'Qc}, \ref{Z'Uc}, \ref{quark:constraint3}, \ref{quark:constraint5}) lead to 
	\begin{eqnarray}
	\psi_{Q+}A\sin(M\pi R)&=&-\frac{y_{U5} H_u}{2}\psi_{U-}^C C^*\cos(M\pi R)\nonumber
	\\
	&=&\left(\frac{y_{U5} H_u}{2}\right)^2\psi_{Q+}A \frac{\cos^2(M\pi R)}{\sin(M\pi R)} \ .
	\end{eqnarray}
Hence, a  consistency condition of quark mass is 
	\begin{eqnarray}
	\tan^2(M\pi R) = \left(\frac{y_{U5} H_u}{2}\right)^2.  \label{eq:mfcondition}
	\end{eqnarray}
The lowest mass eigenvalue gives top quark mass which is not simply proportional to Higgs, 
	\begin{eqnarray} 
	 M_t(H_u) \equiv \frac{1}{\pi R}\arctan\left(\frac{y_{t5} H_u}{2}\right)
	  , \label{app:topmass}
	\end{eqnarray}
and general mass eigenvalues of top quark are 
	\begin{eqnarray}
	M=\frac{n}{R}\pm M_t(H_u)  \quad     (n:\rm integer). 
	\end{eqnarray}
These eigenvalues are used to calculate the effective potential. 

We give the KK expansion for top quark for completeness, 
	\begin{align}
	\Psi_Q
	&= \sum_{n=-\infty}^\infty N_f	t_L^{(n)}(x) \left[	\cos\left(\frac{n}{R}+M_t \right)y
	+\tan(M_t \pi R)\,{\rm sign}(y) \sin\left(\frac{n}{R}+M_t \right)y \right]  \nonumber
	\\
	&\ +\sum_{n=-\infty}^\infty N_f t_R^{(n)}(x)
	\left[-\sin\left(\frac{n}{R}+M_t \right)y
	+\tan(M_t \pi R)\,{\rm sign}(y) \cos\left(\frac{n}{R}+M_t \right)y \right], 
	\\
	\Psi_U^{C} 
	&=
	\sum_{n=-\infty}^\infty N_f t_R^{(n)}(x)  \left[	\cos\left(\frac{n}{R}+M_t \right)y
	+\tan(M_t \pi R)\,{\rm sign}(y) \sin\left(\frac{n}{R}+M_t \right)y \right]\nonumber
	 \\
	 &\ +\sum_{n=-\infty}^\infty N_f t_{L}^{(n)}(x)	
	 \left[-\sin\left(\frac{n}{R}+M_t \right)y
	+\tan(M_t \pi R)\,{\rm sign}(y) \cos\left(\frac{n}{R}+M_t \right)y \right]. 
	\end{align}
where $N_f =\frac{\cos (M_t\pi R)}{(2\pi)^{1/2}}$ is a normalization factor. 

\subsection{Solutions for Squarks}\label{app:squark}
Similarly to the quark case, we solve Klein-Gordon equations for squarks in the bulk respecting properties of $\cal Z$ reflection, 
	\begin{eqnarray}
	\tilde{Q}(y)&=&A_1\cos(M y)+B_1{\rm sign}(y) \sin(M y)   \label{waveQ}\ , \\
	\tilde{U}^{c}(y)&=&C_1{\rm sign}(y)\cos(M y)+D_1 \sin(M y)  \label{waveUc}\ ,\\
	\tilde{U}^*(y)&=&A_2\cos(M y)+B_2{\rm sign}(y) \sin(M y) \label{waveU}\ , \\
	\tilde{Q}^{c*}(y)&=&C_2{\rm sign}(y)\cos(M y)+D_2 \sin(M y) \label{waveQc}\ .
	\end{eqnarray}
These profiles of Eqs.~(\ref{waveQ}-\ref{waveQc}) are valid for an interval of $-2\pi R\leq y\leq2\pi R$.  In principle $\delta(y)$ terms  are allowed for $\tilde{Q}$ and $\tilde{U}$, but we neglect those because we can easily show these should terms vanish by EOMs. Discontinuity at $y=0$ is necessary to take into account Higgs effect  localized on the brane. 
For the translation, squarks have twisted boundary conditions, 
	\begin{eqnarray}
	\left(\begin{array}{c}\tilde{Q}(y+2\pi R) \\  \tilde{Q}^{c*}(y+2\pi R)\end{array}\right)=
	\left(\begin{array}{cc}\cos(2\pi \alpha) & \sin(2\pi \alpha) \\ -\sin(2\pi \alpha) & \cos(2\pi \alpha)\end{array}\right)
	\left(\begin{array}{c}\tilde{Q}(y) \\  \tilde{Q}^{c*}(y)\end{array}\right),  \label{eq:Qalpha}
	\\	
	\left(\begin{array}{c}\tilde{U}(y+2\pi R) \\  \tilde{U}^{c*}(y+2\pi R)\end{array}\right)=
	\left(\begin{array}{cc}\cos(2\pi \alpha) & \sin(2\pi \alpha) \\ -\sin(2\pi \alpha) & \cos(2\pi \alpha)\end{array}\right)
	\left(\begin{array}{c}\tilde{U}(y) \\  \tilde{U}^{c*}(y)\end{array}\right).  \label{eq:Ualpha}
	\end{eqnarray}

We determine the coefficients of the general solutions by EOMs and the above twisted boundary conditions.  Here, only two EOMs of Eqs.~(\ref{EOM1}, \ref{EOM3}) are relevant. We consider integral of Eq.~(\ref{EOM1}) around $y=0$, 
	\begin{align}
	0&=\int_{-\epsilon}^\epsilon \!\!\!dy \ \ \left\{ 
	(M^2+\partial^2_5)\tilde{Q}(y) -y_{U5}^2 H_u^2 \delta(y)\delta(0)\tilde{Q}(y)
	+y_{U5} H_u \delta(y)\partial_5 \tilde{U}^{c}(y)	\right\} \nonumber
	\\
	&=\int_{-\epsilon}^\epsilon \!\!\!dy \  \{M^2  \tilde{Q}(y)\}  + \partial_5 \tilde{Q}(\epsilon) -\partial_5 \tilde{Q}(-\epsilon)
	-y_{U5}^2 H_u^2 \delta(0)\tilde{Q}(0)+y_{U5} H_u \partial_5 \tilde{U}^{c}(0)   \nonumber
	\\
	&=2M B_1 -y_{U5}^2 H_u^2 \delta(0)A_1 +y_{U5} H_u (2\delta(0) C_1 +M D_1) \ . 
	\end{align}
Therefore comparing coefficients leads to two conditions, 
	\begin{eqnarray}
	C_1=\frac{y_{U5} H_u}{2}A_1, \ \ B_1=-\frac{y_{U5} H_u}{2}D_1 \ .   \label{condition1}
	\end{eqnarray}
Similarly, Eq.~(\ref{EOM3}) gives
	\begin{eqnarray}
	C_2=\frac{y_{U5} H_u}{2}A_2, \ \ B_2=-\frac{y_{U5} H_u}{2}D_2 \ . \label{condition2} 
	\end{eqnarray}

Next we apply boundary conditions of the translation.  The condition for $\tilde Q$ of Eq.~\eqref{eq:Qalpha} leads to 
	\begin{align}
	\left(\begin{array}{c}\tilde{Q}(2\pi R-\epsilon) \\  \tilde{Q}^{c*}(2\pi R-\epsilon)\end{array}\right)=&
	\left(\begin{array}{cc}\cos(2\pi \alpha) & \sin(2\pi \alpha) \\ -\sin(2\pi \alpha) & \cos(2\pi \alpha)\end{array}\right)
	\left(\begin{array}{c}\tilde{Q}(-\epsilon) \\  \tilde{Q}^{c*}(-\epsilon)\end{array}\right)
	\\	
	\left(\begin{array}{c}A_1 \cos(2\pi M R)+B_1 \sin(2\pi M R) \\  C_2 \cos(2\pi M R)+D_2 \sin(2\pi M R) \end{array}\right)=&
	\left(\begin{array}{cc}\cos(2\pi \alpha) & \sin(2\pi \alpha) \\ -\sin(2\pi \alpha) & \cos(2\pi \alpha)\end{array}\right)
	\left(\begin{array}{c}A_1\\  -C_2\end{array}\right)\ .
	\end{align}
Eq.~\eqref{eq:Ualpha} leads to a similar equation  exchanging subscripts, $1\leftrightarrow 2$.  
Hence we have, 
	\begin{eqnarray}
	A_1\left[\cos(2\pi M R) -\cos(2\pi \alpha)\right] +B_1\sin(2\pi M R)+C_2\sin(2\pi \alpha)&=&0\ , \label{conditionT1}
	\\
	C_2\left[\cos(2\pi M R) +\cos(2\pi \alpha)\right] +D_2\sin(2\pi M R)+A_1\sin(2\pi \alpha)&=&0\ , \label{conditionT2}
	\\
	A_2\left[\cos(2\pi M R) -\cos(2\pi \alpha)\right] +B_2\sin(2\pi M R)+C_1\sin(2\pi \alpha)&=&0\ , \label{conditionT3}
	\\
	C_1\left[\cos(2\pi M R) +\cos(2\pi \alpha)\right] +D_1\sin(2\pi M R)+A_2\sin(2\pi \alpha)&=&0\ .\label{conditionT4}
	\end{eqnarray}
There are enough conditions of Eqn.(\ref{condition1}, \ref{condition2}, \ref{conditionT1}-\ref{conditionT4}) for 8 coefficients. We solve them, 
	\begin{align}
	&A_1\left[(1+\xi^2)\cos(2\pi M R) -(1-\xi^2)\cos(2\pi \alpha)\right] +2 A_2 \xi \sin(2\pi \alpha)  =0\ , \label{eq:A1}
	\\
	&A_2\left[(1+\xi^2)\cos(2\pi M R) -(1-\xi^2)\cos(2\pi \alpha)\right] +2 A_1 \xi \sin(2\pi \alpha)  =0\ . \label{eq:A2}
	\end{align}
where $\xi \equiv  \frac{y_{U5}H_u}{2}$. Because $A_1=\pm A_2$ is obviously a solution of Eqs~(\ref{eq:A1},  \ref{eq:A2}), a consistency condition  is found to be
	\begin{eqnarray}
	\cos(2\pi M R) &=&\cos(2\pi \alpha \pm 2\theta),    \label{eq:mbcondition}
	\end{eqnarray}
where $\tan\theta\equiv \xi$ (or $\theta =M_t \pi R)$. Hence squark mass eigenvalues are given by 
	\begin{align}
		M=\frac{n+\alpha}{R} \pm M_t \quad (n:\rm integer). 
	\end{align}

For completeness, we determine  coefficients. In a case of $A_1=A_2$,  
	\begin{align}
	&B_1=B_2=C_1=C_2 =A_1\tan(M_t\pi R) , \quad 
	 D_1=D_2=-A_1,
	\end{align}
and in the other case of $A_1=-A_2$,
	\begin{align}
	B_1 =-B_2=-C_1=C_2 =- A_1\tan(M_t\pi R) &\ ,    \quad 
	D_1=-D_2=A_1 . 
	\end{align}
 The KK expansion for squark is
	\begin{align}
	\tilde{Q}&=\sum_{n=-\infty}^\infty \!\!N_b\tilde{t}_{1}^{(n)}(x)
	\left[\cos \left(\frac{n+\alpha}{R}+M_t\right)y
	+\tan(M_t\pi R)\, {\rm sign}(y) \sin \left(\frac{n+\alpha}{R} +\frac{\theta}{\pi R} \right)y\right] 
	\nonumber\\
	&\ +\sum_{n=-\infty}^\infty \!\!N_b\tilde{t}_{2}^{(n)}(x)
	\left[\cos \left(\frac{n+\alpha}{R}-M_t  \right)y
	-\tan(M_t\pi R)\,{\rm sign}(y) \sin \left(\frac{n+\alpha}{R} -M_t  \right)y\right]  ,
	\label{waveQ2} 
	 \\
	\tilde{Q}^{c*}&=\sum_{n=-\infty}^\infty \!\!N_b \tilde{t}_{1}^{(n)}(x)
	\left[-\sin \left(\frac{n+\alpha}{R}+M_t \right)y
	+\tan(M_t\pi R)\, {\rm sign}(y) \cos \left(\frac{n+\alpha}{R} +M_t \right)y\right] 
	\nonumber\\
	&\  -\sum_{n=-\infty}^\infty \!\!N_b \tilde{t}_{2}^{(n)}(x)
	\left[\sin \left(\frac{n+\alpha}{R}-M_t  \right)y
	+\tan(M_t\pi R)\,{\rm sign}(y) \cos \left(\frac{n+\alpha}{R} -M_t  \right)y\right] , 
	 \label{waveQc2}
	 \\
	\tilde{U}^*&=\sum_{n=-\infty}^\infty \!\!N_b \tilde{t}_{1}^{(n)}(x)
	\left[\cos \left(\frac{n+\alpha}{R}+M_t \right)y
	+\tan(M_t\pi R)\, {\rm sign}(y) \sin \left(\frac{n+\alpha}{R} +M_t \right)y\right] 
	\nonumber\\
	&\ -\sum_{n=-\infty}^\infty \!\!N_b \tilde{t}_{2}^{(n)}(x)
	\left[\cos \left(\frac{n+\alpha}{R}-M_t  \right)y
	-\tan(M_t\pi R)\,{\rm sign}(y) \sin \left(\frac{n+\alpha}{R} -M_t  \right)y\right] ,
	 \label{waveU2}
	\\
	\tilde{U}^{c}&=\sum_{n=-\infty}^\infty \!\!N_b \tilde{t}_{1}^{(n)}(x)
	\left[-\sin \left(\frac{n+\alpha}{R}+M_t \right)y
	+\tan(M_t\pi R)\, {\rm sign}(y) \cos \left(\frac{n+\alpha}{R} +M_t \right)y\right] 
	\nonumber\\
	&\  +\sum_{n=-\infty}^\infty \!\!N_b \tilde{t}_{2}^{(n)}(x)
	\left[\sin \left(\frac{n+\alpha}{R}-M_t  \right)y
	+\tan(M_t\pi R)\,{\rm sign}(y) \cos \left(\frac{n+\alpha}{R} -M_t  \right)y\right] , 
	 \label{waveUc2}
	\end{align}
where $N_b=  \frac{\cos (M_t\pi R)}{2\pi^{1/2}}$ is a normalization factor.

\section{Infinite Sum}\label{app:infinitesum}
In order to deal with infinite sums, we replace each element of a sum with a pole in a complex integral. 
For a function that has no singularities on the real $z$ axis, $f(k,z)$, there is a useful relation,
	\begin{eqnarray}
	\sum_{n=-\infty}^{\infty} f(k, n+\alpha)= \sum_{n=-\infty}^{\infty} \oint_{C_n^\alpha}dz \ 
	f(k,z) \frac{\coth[i \pi(z-\alpha)]}{2}, 
	\end{eqnarray}
where the contour $C_n^\alpha$ is a path which rounds about $z=n+\alpha$ with an infinitesimal radius. 
When ${z\to\alpha}$, 
	\begin{eqnarray}
	\frac{\coth[i \pi(z-\alpha)]}{2} = \frac{1}{2i \pi (z-\alpha)} +{\cal O}(z-\alpha) \ , 
	\end{eqnarray}
and the function above is periodic under a transformation of $z \to z+ n\pi$, so each $\oint_{C_n^\alpha}dz$ generates a discrete point of $f(k,z)$. 
\begin{figure}[t]
 \begin{center}
  \includegraphics[width=0.55\linewidth]{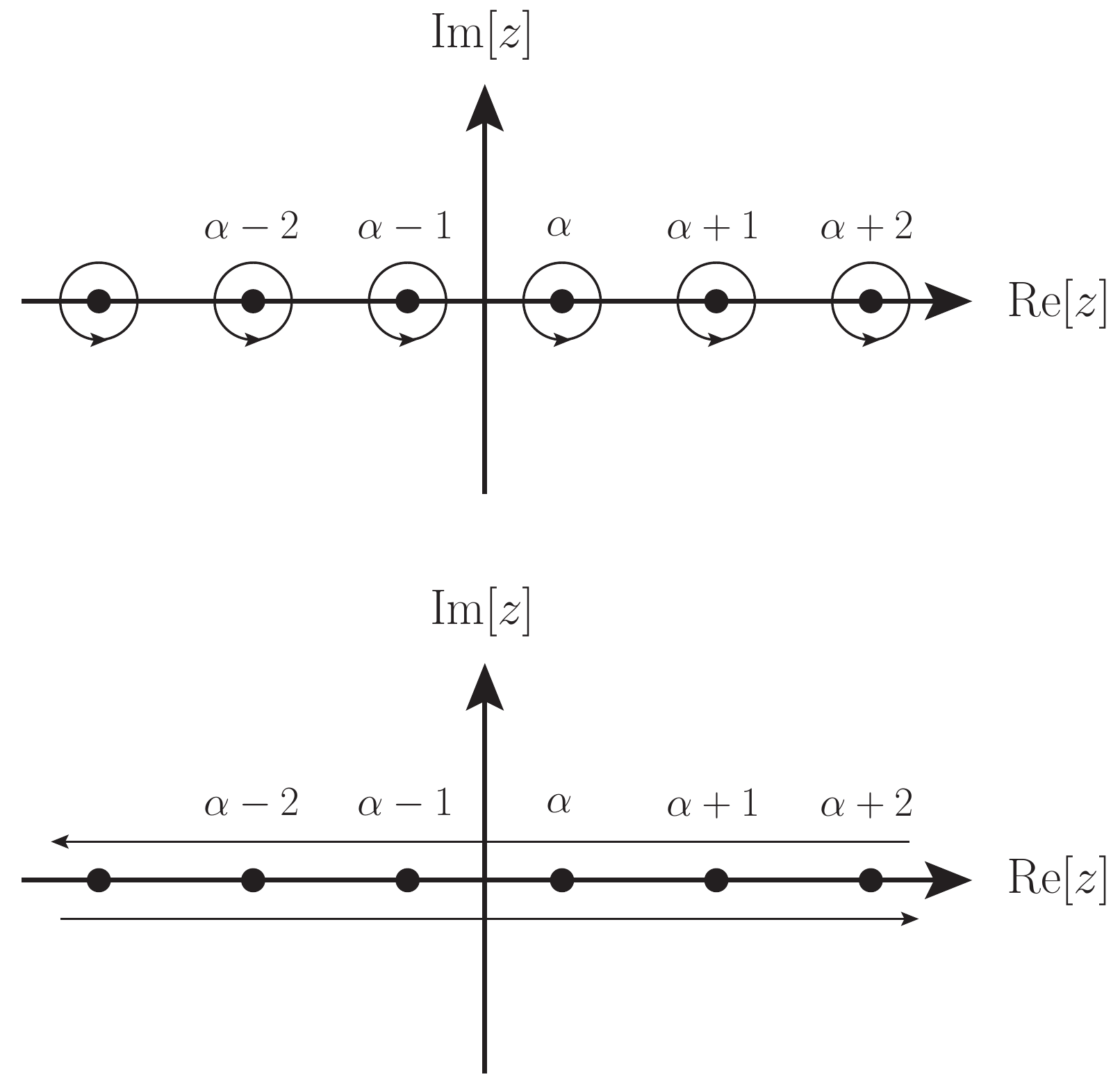}
 \end{center}
 \caption{Integral path of $C_n^\alpha$ in the upper plot. They are combined as shown in the lower plot.}
 \label{fig:path}
\end{figure}
We combine all the contours to paths as in Fig.~\ref{fig:path} and obtain a form with a simple integral,
	\begin{align}
	&\!\!\!\!\sum_{n=-\infty}^{\infty} \oint_{C_n^\alpha}dz \ f(k,z) \frac{\coth[i \pi(z-\alpha)]}{2}
	\nonumber\\&=
	\left(\int_{\infty+i\epsilon}^{-\infty+i\epsilon} dz +\int_{-\infty-i\epsilon}^{\infty-i\epsilon} dz	\right)
	  f(k,z) \frac{\coth[i \pi(z-\alpha)]}{2}
	\nonumber\\&=
	 \int_{-\infty-i\epsilon}^{\infty-i\epsilon} dz	\left\{
	  f(k,z) \frac{\coth[i \pi(z-\alpha)]}{2} -f(k,-z) \frac{\coth[i \pi(-z-\alpha)]}{2}
	 \right\} \nonumber
	\\&=
	 \int_{-\infty}^{\infty} dz	\left\{
	  \frac{f(k,z)+f(k,-z)}{2} \right\}
	\quad +\int_{-\infty-i\epsilon}^{\infty-i\epsilon} dz\left\{
	  \frac{f(k,z)}{e^{2i \pi(z-\alpha)} -1} +\frac{f(k,-z)}{e^{2i \pi(z+\alpha)} -1}
	 \right\}
	 \label{f1}	\ . 
	\end{align}
Here, we used
	\begin{eqnarray}
	\coth(x)=1+\frac{2}{e^{2x}-1}=-\left( 1+\frac{2}{e^{-2x}-1} \right)\ .
	\end{eqnarray}
We are interested in $f(k,z)=1/(k^2+z^2),\ z/(k^2+z^2) $, and hence, for such functions that damps 
for $z\to \pm\infty-i\epsilon$ and can be suppressed by $e^{-2i\pi z}$ for $z \to -i\infty$,  the second expression of Eq.~(\ref{f1}) can enclose the path, referred as to $C_\bigtriangledown$,  in the negative imaginary $z$ plane,
	\begin{align}
	\sum_{n=-\infty}^{\infty} f(k, n+\alpha)
	&=	 \int_{-\infty}^{\infty} dz	\left\{
	  \frac{f(k,z)+f(k,-z)}{2} \right\}
	+\oint_{C_\bigtriangledown} dz\left\{
	  \frac{f(k,z)}{e^{2i \pi(z-\alpha)} -1} +\frac{f(k,-z)}{e^{2i \pi(z+\alpha)} -1}
	 \right\}.  \
	 \label{formula:sum}
	\end{align}
If $f(k,z)$ has poles inside the closed path $C_\bigtriangledown$, the second term on the right-hand side becomes a function of $k$, otherwise it vanishes. In the following section, we will see the cases of, 
	\begin{eqnarray}
	f(k,z)=\frac{1}{k^2+z^2}, \ \frac{z}{k^2+z^2} \ .
	\end{eqnarray}
Using Eq.~(\ref{formula:sum}), formulae of infinite sum are 
	\begin{align}
		\sum_{n=-\infty}^{\infty} \frac{1}{k^2+(n+\alpha)^2} \label{formula:F1}
	=&	\int_{-\infty}^\infty \frac{dz}{k^2+z^2} 
	+\frac{\pi}{k}\left\{ \frac{1}{e^{2 \pi(k-i\alpha)} -1} +\frac{1}{e^{2 \pi(k+i\alpha)} -1} \right\} \ ,
	\\
	\sum_{n=-\infty}^{\infty} \frac{(n+\alpha)}{k^2+(n+\alpha)^2}  	\label{formula:F2} 
	=&
	(-i\pi) \left\{ \frac{1}{e^{2 \pi(k-i\alpha)} -1} -\frac{1}{e^{2 \pi(k+i\alpha)} -1} \right\} \ .
	\end{align}
This  is finite because the momentum dependence is exponentially suppressed in the UV regime. 
The divergent part appear as the first term of Eq.~(\ref{formula:F1}) which has a higher power of UV divergence. However, this divergence is insensitive to supersymmetry breaking parameter, $\alpha$, and then it completely vanishes after combining bosonic and fermionic contribution.  Only finite pieces depend on $\alpha$. 
This is consistent with the non-local nature of supersymmetry breaking by the Scherk-Schwarz mechanism. Because fields notice supersymmetry breaking only when they travel around the extra dimension,  local effects still hold supersymmetric nature leading to absence of UV divergence. 

Finally, we calculate the following sum and integral for the effective potential,
	\begin{align}
	W'(\omega)&=\sum_{n=-\infty}^\infty \int \frac{d^4 k}{(2\pi)^4}  \frac{(n+\omega)}{k^2 +(n+\omega)^2}
	\nonumber\\&=
	\int \frac{d k\ k^3 (2\pi^2)}{(2\pi)^4}(-i\pi) \left\{ \frac{1}{e^{2 \pi(k-i\omega)} -1} -\frac{1}{e^{2 \pi(k+i\omega)} -1} \right\}	\nonumber\\&=
	\frac{-3i}{2(2\pi)^5} [ {\rm Li}_{4}(e^{2\pi i \omega})- {\rm Li}_{4}(e^{-2\pi i \omega})], 
	\end{align}
where
	\begin{align}
	\int_{0}^\infty dk \frac{k^n}{e^{2 \pi(k\mp i\omega)} -1} 	=\frac{n!}{(2\pi)^{n+1}}{\rm Li}_{n+1}(e^{\pm2\pi i \omega})\label{formula:G1}. 
	\end{align}
Here the divergence piece does not appear because  $\omega$ independent terms are already subtracted when $W$ is constructed as in Eq.~\eqref{eq:W}. 

\section{Effect of Higher Order Terms of ${\cal O}(H_u^6 R^{6})$  to Higgs Mass}\label{app:higherterm}
Top quark wave function we solved tells that top  mass is polynomial of $H_u$ in Eq.~\eqref{topmass}. For our Higgs mass calculation, we use an expansion with respected $H_u R$ and take into account up to Higgs quartic terms as in Eq.~\eqref{Huexpand}. Here we show this is good approximation in parameter of our interest, $v\ll 5~{\rm TeV}\lesssim R^{-1}$. 

  \begin{figure*}[t]
 \begin{center}
  \includegraphics[width=0.55\linewidth]{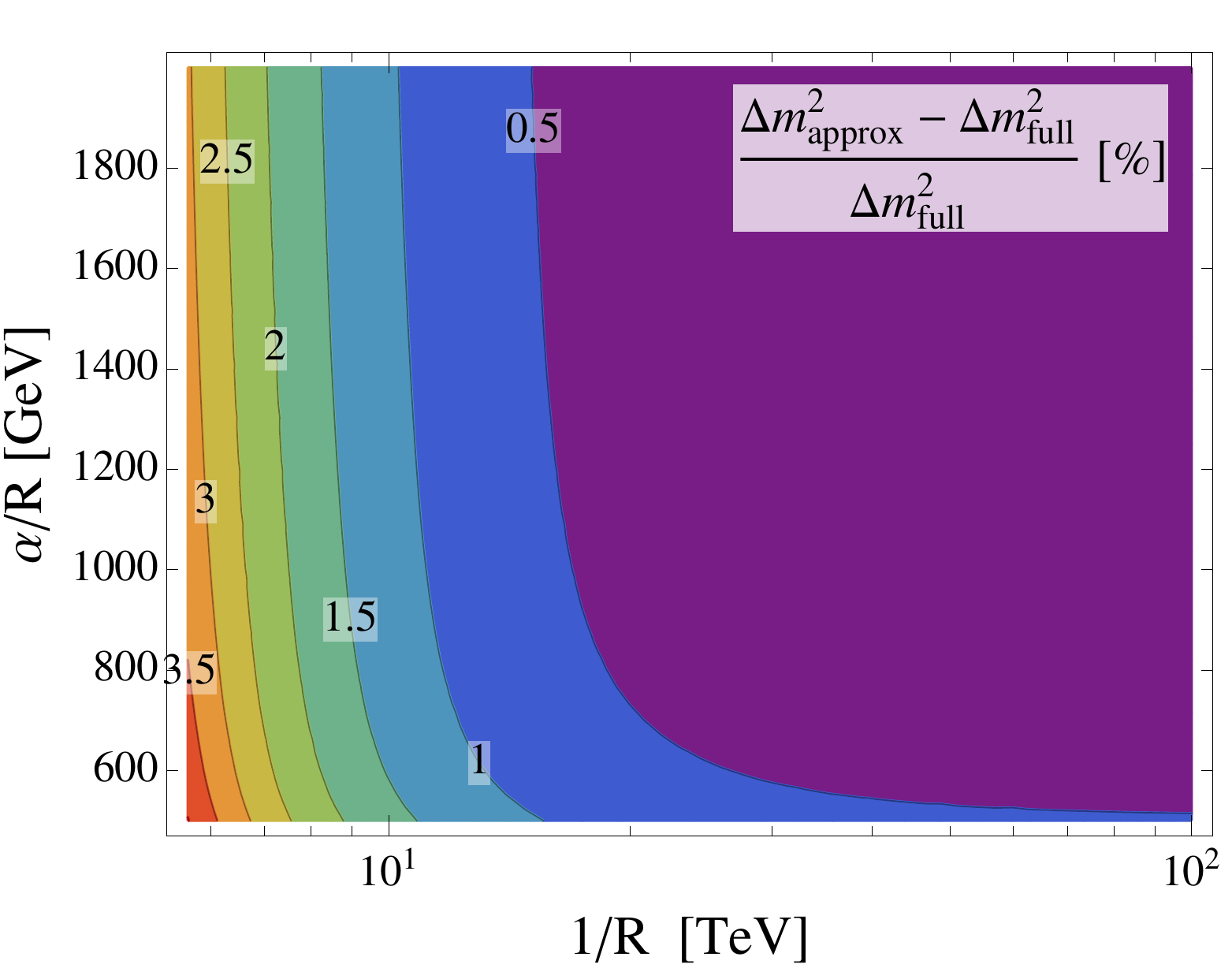}
\vspace{-8pt}
 \end{center}
 \caption{\label{fig:comparison2}
Deviation of $(\Delta m^2_{\rm approx}-\Delta m^2)/\Delta m^2$ in percent. We fix parameters as $v_u=174$ GeV and $y_t=1$ for simplicity.
 }
\end{figure*}

To simplify discussion, we consider the following potential depending on only $H_u$,  
	\begin{align}
	V(H_u)=(\mu^2+m^2_{H_u}) H_u^2 +\frac{g^2_Z}{4}H_u^4+V_t(H_u)
	\end{align}
where $g_Z^2 \equiv {(g^2+g'^2)}/{2}$ and $V_t$ is potential from the radiative correction. 
The second derivative around the VEV gives the Higgs mass, 
	\begin{align}
	m_h^2 \simeq\frac{1}{2}\frac{\partial^2 V}{\partial v_u^2}
	&=(\mu^2+m_{H_u}^2)+\frac{3g_Z^2}{2}v_u^2
	+\frac{1}{2}\frac{\partial^2  V_t}{\partial v_u^2 }
	=g_Z^2 v_u^2
	+\frac{1}{2} \left( \frac{\partial^2  V_t}{\partial v_u^2}
	- \frac{1}{v_u}\frac{\partial  V_t}{\partial v_u}\right) \label{mh2}
	\end{align}
where we used the vacuum condition in the last step. 
 In our estimate we expand the potential with respect to $H_u R$ and  take up to $H_u^4$ terms. 
In order to verify this approximation in parameter of our interest, $v\ll 1/R$, we numerically evaluate the last term of Eq.~\eqref{mh2},
	\begin{align}
	 \Delta m^2 \equiv 
	 \frac{1}{2} \left( \frac{\partial^2  V_t}{\partial v_u^2}
	- \frac{1}{v_u}\frac{\partial  V_t}{\partial v_u}\right)\ .
	\end{align}
In case that the full formula of $V_t$ without any expansion, Eq.~\eqref{eq:Vtfinite}, is adopted, we refer to it as $\Delta m^2_{\rm full}$. On the other hand, we denote $\Delta m^2_{\rm approx}$ is  $\Delta m^2$ using approximated $V_t$ that includes up to $H_u^4$ terms as in Eq.~\eqref{Huexpand}. 
Deviation due to the approximation, $(\Delta m^2_{\rm approx}-\Delta m^2_{\rm full})/\Delta m^2_{\rm full}$, is shown in Fig.~\ref{fig:comparison2}, and it is at most a few percent and is less than percent for $R^{-1}\gtrsim 10$~TeV. This is because the higher order terms are suppressed by $v_u R \ll 1 $. Since this radiative correction contributes to a half of Higgs mass-squared, the Higgs mass changes only by about a quarter of the deviation, $(\Delta m^2_{\rm approx}-\Delta m^2_{\rm full})/\Delta m^2_{\rm full}$.  
 Therefore the approximation of Eq.~\eqref{Huexpand} is valid, and the higher order terms of ${\cal O}(H_u^6 R^{6})$ can be neglected.

\section{Threshold Corrections to Higgs Soft Terms}\label{app:threshold}
Including all the KK tower, we obtain finite results for the Higgs soft terms through 1-loop calculation, 
	\begin{align}
	\delta m_{H_u}^2
	\nonumber=&
	\frac{N_c y_t ^2}{16\pi^2}\frac{3}{\pi^2 R^2}
	\left[ {\rm Li}_3(e^{2\pi i \alpha}) +{\rm Li}_3(e^{-2\pi i \alpha})-2\zeta(3)\right]\\
	&+\frac{\sum_{A=1,2} C_A^h g_A^2}{16\pi^2}\frac{-2}{\pi^2 R^2}
	\left[ {\rm Li}_3(e^{2\pi i \alpha}) +{\rm Li}_3(e^{-2\pi i \alpha})-2\zeta(3)\right]\ , \\
	\delta m_{H_d}^2=&
	\frac{\sum_{A=1,2} C_A^h g_A^2}{16\pi^2}\frac{-2}{\pi^2 R^2}
	\left[ {\rm Li}_3(e^{2\pi i \alpha}) +{\rm Li}_3(e^{-2\pi i \alpha})-2\zeta(3)\right
	]\ , \\
	\delta b
	=&	\frac{i \mu }{16\pi^2}\frac{-N_c  y_t^2 +2 \sum_{A=1,2} C_A^h g_A^2}{\pi R}
	\left[ {\rm Li}_2(e^{2\pi i \alpha}) -{\rm Li}_2(e^{-2\pi i \alpha})\right]\ .
	\end{align}
The Casimir invariants for $SU(2)_L$ and $U(1)_Y$ in $SU(5)$ normalization are $C_{2,SU(2)_L}^h=3/4$ and $C_{2,U(1)_Y}^h=3/20$. The complete diagrammatic calculations are found in Ref.~\cite{Tobioka:2015tua}. 
Since we  match the theory with the MSSM at about compactification scale and solve EWSB conditions, we subtract the MSSM contributions (with the $\overline{\rm DR}$ scheme), 
	\begin{eqnarray}
	\delta{m_{H_u}^2}(Q)&=&\frac{N_cy_t^2}{16\pi^2}\left(\frac{\alpha}{R}\right)^2
	\left\{6\log\left[\frac{Q^2}{(2\pi R )^{-2}}\right]	-16\right\}
	\nonumber\\&&
	+\frac{\sum_{A=1,2} C_A^h g_A^2}{16\pi^2}\left(\frac{\alpha}{R}\right)^2
	\left\{-4\log\left[\frac{Q^2}{(2\pi R )^{-2}}\right]	+8\right\}+{\cal O}(\alpha^4), 
	\\
	\delta{m_{H_d}^2}(Q)&=&
	\frac{\sum_{A=1,2} C_A^h g_A^2}{16\pi^2}\left(\frac{\alpha}{R}\right)^2
	\left\{-4\log\left[\frac{Q^2}{(2\pi R )^{-2}}\right]	+8\right\}+{\cal O}(\alpha^4), 
	\\
	\delta{b}(Q)&=&
	\frac{N_cy_t^2}{16\pi^2}\mu \left(\frac{\alpha}{R}\right)
	\left\{-2\log\left[\frac{Q^2}{(2\pi R )^{-2}}\right]	+4\right\}
	\nonumber\\&&
	+\frac{\sum_{A=1,2} C_A^h g_A^2}{16\pi^2}\mu\left(\frac{\alpha}{R}\right)
	\left\{4\log\left[\frac{Q^2}{(2\pi R )^{-2}}\right]	-4\right\}+{\cal O}(\alpha^3). 
	\end{eqnarray}
We check the IR effect such as $\log\alpha$ terms is certainly cancelled. And  the renormalization scale we choose, $Q_{\rm RG}=\frac{1}{2\pi R}$,
the Higgs soft terms  are 
\begin{eqnarray}
  m_{H_u}^2 
  &=& \left( -\frac{3 y_t^2}{\pi^2} 
    + \frac{3 (g_2^2 + g_1^2/5)}{8\pi^2} \right) 
    \left( \frac{\alpha}{R} \right)^2,
\\[5pt]
  m_{H_d}^2 
  &=& \frac{3 (g_2^2 + g_1^2/5)}{8\pi^2} 
    \left( \frac{\alpha}{R} \right)^2,
\\[5pt]
  b 
  &=& \left( \frac{3 y_t^2}{4\pi^2} 
    - \frac{3 (g_2^2 + g_1^2/5)}{16\pi^2} \right) 
    \mu\frac{\alpha}{R}.
\label{eq:corr-2}
\end{eqnarray}
These results are slightly different from those in Ref.~\cite{Murayama:2012jh}, since Ref.~\cite{Murayama:2012jh} uses  the cutoff regularization  to matched the theories.

\bibliography{SSSBref}
\bibliographystyle{jhep}
\end{document}